# iMER: Iterative Process of Entity Relationship and Business Proces Models Extraction from the Requirements

Muhammad Javed[*], and Yuqing Lin
School of Electrical Engineering and Computing,
The University of Newcastle, NSW, Australia

**Abstract**

**Context:** Extracting conceptual models, e.g., entity relationship model or Business Process model, from software requirement document is an essential task in the software development life cycle. Business process model presents a clear picture of required system's functionality. Operations in business process model together with the data entity consumed, help the software developers to understand the database design and operations to be implemented. Researchers have been aiming at automatic extraction of these artefacts from the requirement document.

**Objective:** In this paper, we present an automated approach to extract the entity relationship and business process models from requirements, which are possibly in different formats such as general requirements, use case specification and user stories. Our approach is based on the efficient natural language processing techniques.

**Method:** It is an **i**terative approach of **M**odels **E**xtraction from the **R**equirements (iMER). iMER has multiple iterations where each iteration is to address a sub-problem. In the first iteration, iMER extracts the data entities and attributes. Second iteration is to find the relationships between data entities, while extracting cardinalities is in the third step. Fourth iteration of iMER is to categorize the attributes as input or output of the operations. Business process model is generated in the fifth iteration, containing the external (actors') and internal (system's) operations.

**Evaluation:** To evaluate the performance and accuracy of iMER, experiments are conducted on various formats of the requirement documents. Additionally, we have also evaluated our approaches using the requirement documents which been modified by shuffling the sentences and by merging with other requirements. Comparative study is also performed. The preliminary results show a noticeable improvement.

**Conclusion:** The iMER is an efficient automated iterative approach that is able to extract the conceptual models from the various formats of requirements.

## 1. Introduction

Software requirements are generally collected in Natural Language (NL) [1]. It is essential for all stakeholders to understand the requirements, but it is challenging due to the vagueness of NL and can leads to disagreement in the later stages of the development [25].

Requirements analysis depends on the feelings and thoughts of an analyst [25]. Requirement analysis is a difficult task due to the NL drawbacks [3]. The major drawbacks of NL include but are not limited to churn, incompleteness, inconsistency, and redundancy [2, 3].

Major portion of the software industry is information systems, being used by all types of organizations [13]. Conceptual model, e.g., Entity Relationship (ER) model plays a vital role in the system specification, analysis, and development [7]. The ER model is first proposed by Chen [16] in 1976, who has explained 11 rules to identify the entities, attributes and relationships from the requirements. The extended ER diagram is presented in [17], by introducing some new concepts on generalization and abstraction. Developing ER diagram from requirements is very often the first step for designing a database system which is an important step of Software Development Life Cycle (SDLC) [6-9].

Abstracting and representing graphically of interaction between the system and end-user make it easier to understand the expected functionality of a software system. Business Process (BP) model demonstrates the overall picture of system's functionality by portraying all the user and system operations, interactions and their sequencing. It details the steps involved to produce the required output from a given input. It acts as a blueprint for the implementation phase.

Manually extraction of the conceptual models is tedious, time consuming and error-prone task [9]. Automated transformation helps to maintain the traceability of the requirements [36]. Many techniques have been proposed to extract the conceptual models from requirements. State of the art Natural Language Processing (NLP) techniques are employed for the extraction of conceptual models.

Accuracy and efficiency are the main challenges for extracting the conceptual models, as stated in [18]. One of the key considerations, during the extraction is how the requirements are presented. Requirements may be written in different formats. For instance, common ways of documenting requirements include but are not limited to, general requirements, Use Case Specification (UCS), user stories, etc. [10]. Furthermore, different templates are being used, depending on the organizational practices and the standards being followed, e.g., IEEE830:2998 or in ISO/IEC/IEEE 29148:2011. Format of the requirements may also depend on the application context, resources, expertise and domain knowledge of the requirement engineers.

---
[*] Corresponding author
E-Mail addresses: m.javed@uon.edu.au, uqing.lin@newcastle.edu.au

Overall, there are little restrictions on the requirements in terms of format and sentence structures. Even if we only focus on the UCSs, engineers are using various extensions to the conventional templates. Requirement in the form of user stories could be quite unpredictable. Indeed, it require extra effort to handle and process the requirements in various format with possible complex sentence structures.

In this paper, we are proposing an automated technique to extract ER and BP models from the requirements in various formats, including general requirements, UCSs and user stories. Our approach doesn't limit to these three formats, we are using these three formats to demonstrate that our approach is able to handle wide range of document format, of cause, some minor adjustments might be necessary.

This iterative approach of models extraction from the requirements is to analyze the Type Dependencies (TDs) of each sentence (w.r.t its structures). BP model produced with all internal and external operations and data to be consumed by the operations. Accuracy and efficiency of the iMER has been demonstrated in our experiments.

In the remainder of this paper, Section 2 contains a detailed literature review. An overview of our proposed approach and the detailed explanation of TDs (used to extract the artifacts) is in Section 3. In Section 4, we presented the case studies. Section 5 presents the evaluation on the accuracy and efficiency of the proposed approach. Finally, conclusion and future works are included in Section 6.

## 2. Literature Review

Many automated and semi-automated efforts have been made to extract the conceptual models from requirements. All of them assumed a specific format of the requirements. A comparative study of these approaches is presented at the end of this section in Tables 1 and 2.

Elbendak et al. [3] proposed a semi-automatic approach to extract the class diagram from UCS based on the basic rules explained in [16]. Meziane and Vadera [7] also tried to extract the ER model from requirements using a semi-automated approach. Firstly, the syntax analysis performed to translate each sentence into a Logical Structure form Language (LSL), i.e. *determiner (Base, Focus)* where Base is the noun phrase and Focus is the verb phrase. Nouns as entities and verbs as relations are identified from these translated sentences. Degree of the relationships is identified from the quantifiers, i.e., 'the', 'a / an' and other rules based on the forms of verb. We noticed that nouns, verbs, adjectives, and pronouns were represented in logical form but there is no clear interpretation of pronouns.

Btoush and Hammad [9] presented an approach to extract the ER model. The first step is sentence segmentation. Sentences are separated by considering the period (.) sign and then all punctuation signs were deleted. After tokenization and POS-tagging, words were connected as chunks to get the noun and verb phrases. Then parse tree of each sentence is generated using the memory based shallow parser. According to the rules discussed, nouns and gerund are the entities, transitive verbs, with propositions and indicating words are the relationships between entities. Adverb and object in (subject + possessive verb + adjective + objective) format indicate the primary key. This tool can only process simple sentences.

Overmyer et al. [19] proposed a framework which requires domain exports to manually selecte the words based on their frequency in the text. The words then highlighted with different colours to represent the classes, attributes and operations. Omar et al. [20] tried to extract the ER model by assigning weight to the potential candidates of entity. Human involvement is required to attach attributes with entities and to find the relationships between entities.

Harmain and Gaizauskas [21] proposed an approach to generate class diagram from the requirements. Two versions of the CM-Builder are discussed here. CM-Builder 1 is a semi-automatic tool with eight modules. The first five modules are for pre-processing of text, parsing is to generate the syntactic tree (based on grammatical rules) and semantic representation of every sentence. Then noun phrase frequency analysis module is to generate the candidate classes and attributes from the syntactic trees and their frequency from the text. Verb phrase filter module is to get verb phrases as candidate associations. OOA workbench module is an interface for the users to select the classes, attributes, and their relationships from the lists. CM-Builder 2 is an automated version to extract the class diagram. OO analysis module is responsible for the generation of class diagram. Nouns with the higher frequency are the potential class candidates. Non-copular verbs are the relationships. Possessive relationships, adjectives and verbs e.g., 'to', 'have', 'denote', and 'identify' in the sentences are analyzed to identify the potential attributes. Relationship between related objects and subjects are defined by an attached propositional phrase. Multiplicity of the association is identified from the determiners and denoted by 1(for exactly one), * (for many), and N (for specific number). Once the model generated, analyst can refine it.

The approach of Uduwela and Wijayarathna [22] is to extract the normalized database by identifying the primary key, foreign key and relationships between the data entities from a structured format of requirements (forms). Human involvement required to confirm the selection of entities.

Lilac [23] applied syntactic analysis on requirements by using the context-free grammar rules to collect the words as entities or attributes. Vemuri et al. [24] processed the requirements to collect the actors and use cases using a pre-trained classifier.

Yue et al. [26] proposed an automated approach called aToucan to extract the UML models (class, sequence and activity diagrams) from UCS. This technique process only simple sentences. Multiple packages are included in the metamodel of aToucan. Sentence semantics package is for the classification of sentences into condition or action. A sentence classified as action if it is an input from the primary actor, validating the data, system alter its state, system generate an output, or system sends output to a secondary actor. Sentence structure package is to identify the natural language concepts from the sentences, i.e., object, subject, verb, and noun. It is subdivided into three packages; sentence, phrase, and POS. Sentence package is to get the subject. Phrase package is to find the noun phrases, verb phrases, adjective phrases, adverb phrases, and prepositional phrases from the simple sentences. While POS package is to label the seven part of speech tags of a sentence (Noun, Verb, Preposition, Adverb, Adjective, Conjunction, and Determiner). Set of rules were defined for each model to extract artifacts independently.

Ambriola and Gervasi [27] presented a tool called CIRCE to convert the natural language requirements into models. The author tried to extract static models (ER or class diagrams) and dynamic models (finite state automata or event-condition-action rules) separately. Parse trees obtained from the requirements are converted to tuples and then enhanced using the extensional knowledge about the basic structure of software. Bajwa et al. [28] proposed a tool to generate the class diagram from the requirements. After POS tagging, classes and attributes are extracted from the set of nouns.

Lucassen et al. [2] presented an automated technique to extract the conceptual model from user stories. They followed the Grimm model to combine tools (AQUSA, Visual Narrator, and Interactive Narrator) to get the required output. AQUSA is a tool to identify the defects in the user stories using NLP technique. Then software engineers correct the identified textual and structural errors. Visual Narrator is to extract the artifacts, based on 11 heuristics. Matrix component is to remove the stopping words and to assign the frequency-based weights. Term-by-user-stories matrix is the output of this module. Constructor module is, based on heuristics, to generate the conceptual model by processing the weighted user stories. The algorithm is to extract the elements by splitting the user stories into role, mean, and end. If a story is not dividable then discard it and moves to the next. SpaCy is used for the pre-processing (tokenizing, POS tagging, and dependency generating) of user stories. In Visual Narrator, the nouns are identified and added to the list of entities. Relationships between entities are identified from the associated verbs, e.g., verbs related to the subject and objects of a story are considered as relationships. Interactive Narrator is responsible to present the output of Visual Narrator in an interactive view.

Omer and Wilson [6] tried to extract database design form user requirements using a natural language processing tool, by considering only subject and object of sentences. Stanford CoreNLP 3.3.1 is used for getting nouns, adjectives, verbs, subjects, and objects from the text. According to the steps defined in algorithms, direct subject, direct object, and passive object are the tables and verbs are the relationships between them. Frequency of each table is calculated to finalize the list. However, we find that the attributes extraction criteria is not clearly defined by the authors.

Thakur and Gupta [12] presented an automated approach to extract the class diagram from use cases. They used Stanford NL parser API for the pre-processing of text that involve POS tagging and TDs generation. A language model defined to interpret the pattern of simple and certain types of complex sentences. This language model is based on twenty-five verb patterns (25 ways of using a verb phrase in a sentence). Authors defined rules to recognize the sentence structure by using these verb patterns. Then they applied transformation rules on TDs to extract the classes, attributes, operations, and relationships between classes. In this technique, TDs analysis is based on the type of the sentences.

Arora et al. [15] proposed an automated method to extract domain model from unrestricted natural language requirements and represented in the form of classes. The pre-processing of sentences involved parse tree generation, stop word removal, and lemmatization. Nouns and verbs are separated and TDs are generated. According to the approach, TDs with nouns and verbs are the main source to identify the artefacts. The algorithm takes a list of atomic nouns, verbs, and dependencies as input and apply rules to extract the model. Noun is a candidate concept if it appears in the source or target of a dependency. Verb in subjective or objective part of sentence is the association relation. Generalization is identified from the adjective modifier. Cardinalities are based on the quantifiers appeared with concepts. Attributes were identified manually.

Sagar and Abirami [25] presented a tool to create a conceptual model from the functional requirements except the negative sentences. Pre-processing and syntactic analysis of sentences are the initial steps. They defined some rules to extract the artefacts from TDs. According to these rules, Nouns appeared as subject or object and gerund are the candidate classes. Adjective and noun followed by keywords (identified by, recognized by) are the attributes. Intransitive verb with adverb, possessive apostrophe, and "of" construction are also the indicators of attributes. Relationships identified from the different roles of verbs (transitive, verb with proposition, and appearance in some specified sentence structures). Then these relationships are classified as association, aggregation, composition, or inheritance. Aggregation identified from the patterns 'is made of', 'is part of', 'contain', 'consist', and 'comprises'. Inheritance is based on copula verb, while no clear rules for association and composition relationships are defined.

Karaa et al. [30] presented an automated approach to generate class diagram from the functional requirements. Tokenization, POS tagging, and syntactic parsing are the pre-processing steps. Authors presented rules (regular expression format) to extract the elements of class diagram from TDs. However, there is no discussion about the TDs. This approach also processed the simple sentence formats.

Most techniques analyzed the TDs to extract the artefacts. TDs represents the syntactic relationships of words within a sentence [32]. Table 1 illustrates the TDs processed by the researchers in existing techniques.

**Table 1** TDs considered by the researchers to extract the artefacts

| Proposed by | Type Dependency |
|---|---|
| Lucassen et al. [2] | nsubj (A, B), dobj(A,B), pobj(A,B), nn(A,B), amod(A,B) |
| Omer and Wilson [6] | dobj(A,B), dsubj(A,B), nsubjpass(A,B), nobjpass(A,B), compound(A, B) |
| Thakur & Gupta [12] | nsubj (A, B), nsubjpass(A,B), dobj(A,B), pobj(A,B), iobj(A,B), nn(A,B), amod(A,B), xcomp(A,B), prep(A,"in"), prep(A,"of"), poss(A, B), aux(A,B), num(A,B), neg(A,B), cop(A,B), advmod(A,B), complm(A,B), infmod(A,B), partmod(A,B) |
| Arora et al. [15] | nsubj (A, B), dobj(A,B), amod(A,B), ref_to(A,B), rcmod(A,B), ccomp(A,B), xcomp(A,B), vmod(A,B). |
| Sagar & Abirami [25] | nsubj (A, B), dobj(A,B), amod(A, B), advmod(A, B), nmod:of(A,B) |

In these TDs, "A" and "B" are representing the words of a sentence. In these approaches, only a few TDs are processed due to the assumption of specific format of requirements and restricted sentence structures. Given that limited TDs are considered, most of the existing approaches could not handle more complex structure.

**Table 2** Comparison of domain model extraction approaches w.r.t the input format

| Approaches | General Requirement | Use Cases | | User Stories |
|---|---|---|---|---|
| | | Restricted Template | Unrestricted Template | |
| **Semi-Automated Approaches** | | | | |
| [3] | ✓ | ✓ | | |
| [6] | ✓ | | | |
| [7] | ✓ | | | |
| [19] | ✓ | | | |
| [20] | ✓ | | | |
| [22] | ✓ | | | |
| **Automated Approaches** | | | | |
| [9] | ✓ | | | |
| [21] | ✓ | | | |
| [23] | ✓ | | | |
| [23] | ✓ | | | |
| [24] | ✓ | | | |
| [25] | ✓ | | | |
| [26] | | ✓ | | |
| [27] | ✓ | | | |
| [28] | ✓ | | | |
| [30] | ✓ | | | |
| [12] | | ✓ | | |
| [15] | ✓ | | | |
| [2] | | | | ✓ |
| **iMER** | ✓ | | ✓ | ✓ |

Table 2 contains the formats of NL requirements processed by the researchers. We are proposing an automated technique to process all the three formats of requirements.

Studies on the existing techniques of model extraction shows that the results are unsatisfactory [31]. A few efforts have been made to generate the BP model but to the best of our knowledge, no effort has been made to depict it along with the data involved in the process, such a combination confirms the required functionality of the system.

## 3. Proposed Approach

In this section, we are presenting our extraction process. A tool (called iMER) has been developed in Visual C# to extract the ER and BP models from requirements. Stanford CoreNLP 3.8 API is used for the pre-processing and syntactic analysis of sentences in the requirement document. An overview of steps involved for the extracting process is illustrated in Fig. 1.

Summarized steps of the proposed approach are as follows:
Step 1: for the requirements format such as UCS where requirement break into sections and have reference to each other, first step is the sentence sequencing.
Step 2: Analysis of the sentences. This requires:
   a. Pre-processing
   b. Syntactic Analysis.
Step 3: Analysis of TDs to extract the ER model:
   c. Entities and attributes Extraction
   d. Relationships Extraction
   e. Cardinalities & Modality Extraction
Step 4: Categorization of Attributes
Step 5: BP model extraction

### 3.1. Step 1: Sentence Sequencing

We are using UCS as an example, this step is applied to any document formats where requirements might be splitted into sections or modules, with references between sections.

UCSs contains functional requirements with a sequence of actions performed by the actors and system [12]. Commonly, a UCS contains the sections such as use case name, description, pre-condition, post-condition, actors and main flow. However, alternate flow and exceptions are often presented differently. Different formats of referencing the sentences in alternate or extension sections are being used [38]. For instance, main flow could contain the references to the flows in extension / alternate section at the end of sentences, or the referred sentences in other sections could have the same sentence number as source in the main flow section. It makes a challenging task because an accurate sentence sequencing, referencing format need to be detected. iMER will scan the whole UCSs document, inserting the refered sentences of alternative or extension sections into the main flow and this is what we call *Sentence Sequencing*.

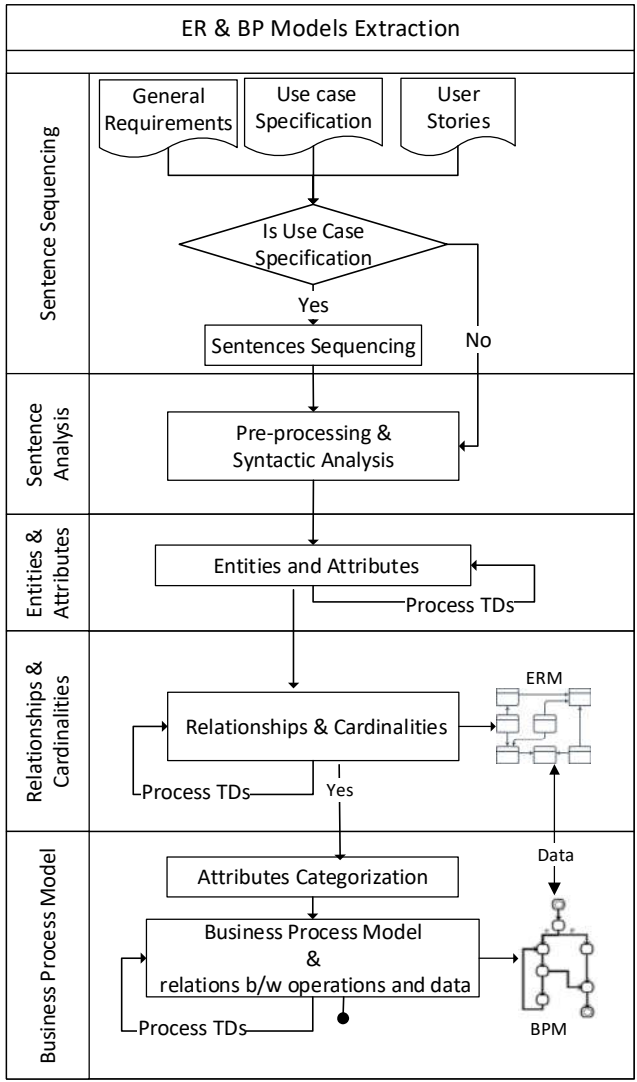

**Fig. 1** Flow chart of the Proposed Approach,

## 3.2. Step 2: Sentence Analysis

### 3.2.1. Pre-Processing

Sentence analysis is the basic and fundamental step of NLP. This step of iMER is for the pre-processing of sentences that involved tokenizing (splitting sentences), lemmatizing (converting each word to its basic form, e.g., played to play) and Part of Speech (POS) tagging (marking the words with part of speech tags as a prefix). This is achieved using the Stanford CoreNLP 3.8 API. Our assumption here is: each sentence should end with a full stop / period sign (.), and should not contain a period or hyphen (-) character within the sentences, for an accurate processing. Stanford parser generates POS-tags in a tree like structure. Following is an example of the parse tree.

*Example:* "A language tape has a title language and level."
(ROOT
 (S
  (NP (DT a) (NN language) (NN tape))
  (VP (VBZ has)
   (NP
    (NP (DT a) (NN title) (NN language))
    (CC and)
    (NP (NN level))))
  (. .)))

In this parse tree, each word of the sentence has a prefix, representing the POS tag. For example, *NN* (noun) is the prefixes of "*language*". Further details on POS-tagging can be found in [39].

**Fig. 2** Visual representation of POS –Tags using Stanford CoreNLP web API[1]

### 3.2.2. Syntactic Analysis

TDs represents the syntactic relationships of words within a sentence [32]. Each TD contains a pair of words based on their grammatical relation. Following are the TDs generated from the sentence used in our previous example in Fig. 2. The first TD is *root* with a false word "ROOT-0" and a verb "has". The second TD is *det* for the relationship of a noun with its determiner. Number with each word in TDs, represents its sequence in a sentence. A detailed explanation of TDs can be found in [32].

*root(ROOT-0, has-4)*
*det(tape-3, a-1)*
*compound(tape-3, language-2)*
*nsubj(has-4, tape-3)*
*det(language-7, a-5)*
*compound(language-7, title-6)*
*dobj(has-4, language-7)*
*cc(language-7, and-8)*
*dobj(has-4, level-9)*
*conj:and(language-7, level-9)*
*punct(has-4, .-10)*

**Fig. 3** Visual representation of TDs using Stanford CoreNLP web API

## 3.3. Step 3: Analysis of Type Dependencies for ER model

### 3.3.1. Entities and Attributes Extraction

In this subsection, we are presenting how iMER extracted the potential entities and attributes by applying the Type Dependency based Rules (TDRs). In these rules, *PrevTD* represents the Previous TD while *nextTD* is for next TD with respect to the current TD under consideration. *Basic_Attrib* represents the basic forms of attributes, e.g., name, number, type, address, level, date, time, etc.

---
[1] http://nlp.stanford.edu:8080/corenlp/process

TDRs defined by analyzing the dependencies independently & w.r.t its co-occurrence TDs. To make the rules easy to understand, we presented them in Antecedent-Consequent format [12]. Rules used by iMER to extract the entities and attributes are described in Appendix A.1 (in the algorithmic format).

From the existing work, it is commonly agreed that nouns, especially common nouns [12, 16, 17, 34, 35] and Gerund [16, 35] are the potential entities while attributes appear as noun, adjective [7, 16, 17, 21, 34, 35], possessive apostrophe [21, 25] or with indicators (e.g., number, name, date etc.) [35]. It is worth mentioning here that not all nouns are the candidate for entities or attributes, for example, database, system, company, record, etc. [35].

TDs are processed to extract the components of artefacts. In here, two-way iterations (forward and backward) are used. The first iteration is to find the entities and attributes by applying the TDRs. Once the entities and attributes are extracted in the forward iteration, iMER traces back to complete the names of artefacts in case of compound phrases.

### 3.3.1.1. Subject

Nouns appearing in the subjective part of a sentence are the potential candidate for entity [28, 35]. In [2, 12, 15, 25, 30] authors tried to process *nsubj* TD to extract the entity, while [12] and [30] also processed *nsubjpass* to extract the entities.

While processing the sentences of different structures, it is noticed that the subject of the sentence might also contain the attributes. In TDs, the subjective part of the sentence is represented by *nsubj(verb, noun)*. We processed the subjective dependencies based on the type of sentences: i.e. *nsubj*: nominal subject, *nsubjpass*: passive nominal subject, *xsubj*: controlling subject, *csubj*: clausal subject, *csubjpass*: clausal passive subject of the sentence. It is suggested that *nsubj* and *nsubjpass* contain entities or attributes.

*Dependency: nsubj(A, B) & nsubjpass(A, B)*

If B part of these TDs is not a member of the basic attributes set, then it is a potential entity (*TDR1 & TDR2 in Appendix A*).
*Example:* "The customer cancels the reservation."
From this example along with other TDs, we get *nsub(cancels-3, Customer-2)* which indicating "*Customer*" as an entity.
*Example:* "An offer is selected by the customer."
From *nsubjpass(selected-4, offer-2)*, "*offer*" is extracted as entity.
*Example:* "ID and password are the basic information for login."
TDs *nsubj (information-7, ID-1)* and *nsubj (information-7, password-3)* suggest the attributes "*ID & password*"

Compound nouns are represented by *compound(A, B)*. If it appears just before the *nsubj* or *nsubjpass,* then the component will be extracted by combining the both A and B of compound TD.
*Example:* "A Credit card has an expiry date."
From TDs *compound(card-3, credit-2)* and *nsubj(has-4, card-3)*, "*Credit Card*" entity is extracted.
*Example:* "Expiry date of cash card is entered by the customer"
From TDs *compound(date-2, expiry-1), nsubjpass(entered-6, date-2)*, attribute "*Expiry Date*" can be generated.

### 3.3.1.2. Object

Object of the sentence contains entities [2, 12, 15, 25, 30]. Normally it is hinted by the TD *dobj(Verb, Noun)*. In [15, 25, 30], the authors processed *dobj* to extract the entities, while in [2, 12] the authors also considered *pobj* TD.

*Dependencies: dobj (A, B), iobj(A, B) & pobj(A, B)*

iMER process objective decencies (*dobj*: direct object, *iobj*: indirect object and *pobj*: object of a preposition) to extract the artefacts. If neither B part contains the basic attributes, nor the operation in A gives the sense of inputting information (e.g., enter, inputted, save, add, has) and nor the previous TD is *amod* (adjectival modifier), then it is an entity. If *compound* appears before any of the object TDs, then entity will be extracted by combining the nouns in *compound* (*TDR3 in Appendix A*).
*Example:* "A customer cancelled the reservation."
From *dobj(cancelled-3, reservation-5),* entity "Reservation" is generated.
*Example:* "The customer selected credit card for the payment."
*compound(card-5, credit-4), dobj(selected-3, card-5)* suggested the entity "Credit Card".

If B part of the object TD contains basic attribute and the previous TD is *amod* or *compound,* then the attribute will be extracted by combining the A and B parts of previous TDs (*TDR4 & TDR5 in Appendix*).
*Example:* "A customer enters phone number"
*compound(number-5, phone -4), dobj(enters-3, number-5)* gives "*Phone Number*" as an attribute.
Example: "A customer enters first name, last name and address"
From *amod ( name-5 , first-4 ), dobj ( enters-3 , name-5 ), amod ( name-8 , last-7 ), dobj ( enters-3 , name-8 ), dobj ( enters-3 , address-10 )*, attributes "*first name, last name and address*" are extracted.

### 3.3.1.3. Prepositions

Prepositions represent the relationship between the nouns or pronouns and other words in the sentence.

*Dependency: nmod:of(A,B)*

In [12] authors considered A as attribute and B an entity for all cases. While we found that any part of *nmod:of* could either be an enitty or attrbute, dependeing on the sentence structure (*TDR6 in Appendix A*).
*Example:* "Visitor selected the type of an event."
*nmod:of(type-3, event-5)* implies attribute "*Type*" and "*Event*" as an entity.
*Example:* "Card of the customer has an expiry date"
Based on *nmod:of(card-1, customer-4)*, "*Card*" and "*Customer*" both are entities.
*Example:* "Visitor entered the date of birth."
*nmod:of(date-3, birth-5)* indicates "*Date of birth*" as an attribute.

*Dependency: nmod:in(A,B)*

In this TD, A might be an attribute and B is an entity (*TDR7 in Appendix A*). In [12], authors presented the same concept.
Example: "The system validates that a customer has enough funds in the account."
Based on *nmod:in(funds-9, account-12)*, "*Fund*" is an attribute and "*Account*" is the entity.

*Dependencies: nmod:to(A,B), nmod:for(A,B), nmod:from(A,B), nmod:as(A,B)*
In these TDs, B will be the potential candidate for entity (*TDR8 in Appendix A*).
*Example:* "The system displays a price to the customer."
From *nmod:to(displays-3, customer-7)*, "*Customer*" is the entity extracted.
*Example:* "system starts displaying video feed for the coordinator."
From *nmod:for(displaying-3, coordinator-7)*, entity "*Coordinator*" is extracted.
*Example:* "Information does not match received from the witness."
*nmod:from(received-5, witness-7)* suggests the entity "*witness*".
Example: "As a visitor, I can create a new account."
*nmod:as(create-7, visitor-3)* suggests "*Visitor*" entity.

*Dependencies: nmod:by(A,B), nmod:agent(A,B), nmod:with(A,B)*
Agent indicates someone or something that performs an operation on the subject [33]. If B part of TDs belong to the set of the basic attributes, then it will be an attribute, otherwise it might be an entity (*TDR9 in Appendix A*).
*Example:* "A branch is uniquely identified by the branch_number."
*nmod:agent(identified-5, branch_number-7)* implies "*branch_number*" is an attribute.
*Example:* "Name and address are entered by the customer"
*nmod:by(entered-4, customer-7)* has entity "*Customer*".

### 3.3.1.4. Appositional modifier

The possessive form is to show relationship between the nouns.

*Dependency: nmod:poss(A,B)*
If A is an attribute, then B might be an entity. In [12] possessive nouns are considered w.r.t to apostrophe. In our approach, along with the possessive apostrophes, we also processed possessive determiners and possessive (*TDR10 in Appendix A*).
*Example:* "The administrator enters customer's address."
In *nmod:poss(address-6, customer-4)*, "*Customer*" is an entity while "address" is an attribute.
*Example:* "The witness provided his name."
*nmod:poss(name-5, his-4)* suggest "*name*" as an attribute.

### 3.3.1.5. Adjectival modifier

An adjectival modifier is to modify the meaning of a noun/noun phrase [32].

*Dependencies: amod(A,B)*
If B is adjective and A is not a basic attribute, then A might be an entity or by combining A and B, an attribute could be generated (TDR11 in Appendix A). While in [12] the authors considered A as an entity and B as an attribute in all cases.
*Example:* "System assign the initial level of emergency."
*amod(level-5, initial-4)* implies "*initial level*" is an attribute.
*Example:* "Coordinator determines that the witness is calling a fake crisis." *amod(crisis-11, fake-10)* suggests "*Crisis*" is an Entity.

### 3.3.1.6. Compound Noun

*Dependency: compound(A,B)*
If this dependency does not appear before the subject or object TDs, then it is considered independently. If any word of the TD is representing the basic attribute, then the other might be an entity, otherwise entity is extracted by combining both words (*TDR12 in Appendix*).
*Example:* "The coordinator provides information (witness ID, first name, last name, phone number, and address)."
*compound(ID-7, witness-6)* suggests "*Witness ID*" attribute and "Witness" entity while *compound(number-16, phone-15)* indicates "*Phone Number*" attribute.
*Example:* "The customer paid by credit card."
In *compound(card-6, credit-5)* has "*credit card*" as entity.

### 3.3.1.7. Conjunction

A conjunction illustrates the relationship between two elements [32].

*Dependency: conj:and(A,B) , conj:or(A,B)*
A & B might be the potential candidate for attributes (*TDR13 in Appendix A*).
*Example:* "A customer enters ID and password to login."

In *Conj:and(ID-4, password-6)*, both *ID* and *password* are the attributes.

### 3.3.1.8. Gerund

Gerund appears as a noun in different TDs [32]. Hence, iMER also processed it.
*Example:* "System display the booking dates."
With *compound (dates-5, booking-4)*, "*Booking*" is taken as an entity.

### 3.3.1.9. Pronoun

A pronoun represents the immediate actor/entity. Therefore, iMER replaces pronouns with their anaphors, i.e., the nouns to which they refer.
*Example:* "User selects login option. He enters ID and password."
In this example, pronoun in second sentence *nsubj(enters-2, He-1)* refers to the subject noun of first sentence "user" in *nsubj(selects-2, user-1)*. Hence, "He" will be replaced with "*User*".
*Example:* "As a visitor, I can create a new account."
In this sentence, TD *nmod:as(create-7, visitor-3)* contains entity and TD *nsubj(create-7, I-5)* contains pronoun. So, pronoun "I" will be replaced with "visitor".

### 3.3.2. Relationships Extraction

A verb, especially a transitive verb, represents the relationships between the entities [16, 17, 21, 27, 34]. Verbs followed by the prepositions are also indications of relationships [35].

In this iteration, iMER reuses the TDs to find the relationships. However, we only consider those TDs that have the entities in it, thus make the iMER more efficient. The rules followed by iMER to generate the relationships are listed in the annexure A.2. In these rules, "E" represents the entities extracted in the previous iteration.

TDs having entity in one part and verb in second are considered. If two dependencies of a sentence have a common verb, then the verb is considered as a relationship between the participating entities (*TDR14 & TDR15 in Appendix A*).
*Example*: "The administrator manages branches."
From *nsub(manages-3, Administrator-2) & dobj(manages-3, branches-4)*, we have "*Administrator (manages) Branches*".

Entities with preposition "of" are connected by the relationship "has" (*TDR16 & TDR17 in Appendix A*).
Example: "Card of the customer has an expiry date"
*nmod:of(card-1, customer-4)* implies "*Customer (has) card*".

iMER attaches the prepositions ("to", "in", "for", and "from") with the main verb of relationships (*TDR18 to TDR22 in Appendix A*). However, in case of comparative modifier "as", only the verb is considered as a relationship (*TDR23 in Appendix A*).
*Example:* "A customer adds items to the cart."

From TDs *nsubj(adds-2, customer-2)*, *dobj(adds-2, items-4)* and *nmod:to(adds-3, cart-7)*, relationships "Customer (adds) items", "items (adds to) cart" and "Customer (adds to) cart" are extracted.

In UCSs, it might be possible that there is no direct relationship between the entities, i.e., entities do not appear in the same sentence. In this case, iMER will find a relationship from the flow of data.
*Example* (sentences from a use case with TDs)**:**
1. "A customer selects the date."
   root ( ROOT-0 , selects-3 ), det ( customer-2 , A-1 ), nsubj ( selects-3 , customer-2 ), det ( date-5 , the-4 ), dobj ( selects-3 , date-5 )
2. "System displays the available booking dates."
   root(ROOT-0, displays-2), nsubj(displays-2, system-1), det(dates-6, the-3), amod(dates-6, available-4), compound(dates-6, booking-5), dobj(displays-2, dates-6), punct(displays-2, .-7).

In this example, "*Customer*" and "*Booking*" are not in the same sentence, but considering the flow of data (i.e., "*Customer*" is selecting a "*date*" in the first sentence that belongs to "*booking*" in the next) iMER generated relationship "*Customer (selects) Booking*."

### 3.3.3. Cardinalities and Modality Extraction

Cardinality signifies how many instances of one entity can be associated with the instance of other entity. Cardinalities can be extracted by tracking the special indicators e.g., "many", "more", "each", "all", and "every", similar approaches used in [21, 35]. This iteration of iMER is to extract the cardinalities of entities. Only those TDs that have the entities are processed and this made the iMER more efficient. Rules related to the cardinality extraction are listed in the annexure A.3.

### 3.3.3.1. Adjective modifier

Adjective modifier (*amod*) relation might contain the cardinality. If the adjective part of the dependency contains any of the keyword (many, some, all, more, every, first or last), then the cardinality will be N (*TDR24 in Appendix*).
*Example*: "A store has many branches."
*amod(branches-5, many-4)* suggests the cardinality '*N*'.

### 3.3.3.2. Number modifier

Number modifier (*nummod*) dependencies help to find the modality or cardinality. If keywords "at least" or "minimum" are the prefix of a number, then it is the modality. If the number has a prefix "at most", "limit", "maximum" or "no more than", then this number is the cardinality (*TDR25 in Appendix*).
*Example:* "Branch must be managed by at most 1 manager."
*nummod(manager-10, 1-9)* represents the cardinality "1".

### 3.3.3.3. Determiner

Determiner (*det*) represents the cardinality of a subjective entity. If any of the keyword "many", "some", "each",

"more", "all" appear as a determiner of an entity then the cardinality is N. If "a/an" appears in front of an entity, the cardinality is one (*TDR26 in Appendix*).

*Example:* "Each product has an expiry date."
*det (product-2, Each-1)* illustrate cardinality "1".

POST-tags, as a prefix with entities can help to extract the cardinality. For instance, if the entity is marked with NNS (Plural Noun), it might represent the N cardinality while entities marked with NN and NP should be the cardinality "1".

### 3.4. Step 4: Categorization of Attributes

This iteration is to classify the attributes as inputs or outputs w.r.t the required functionality. iMER process the TDs containing operations for the separation of inputs and outputs.

In simple sentences, the TDs representing subject and object are sufficient to determine the categorization of data. However, in complex structured sentences, this is more challenging. Therefore, along with the TDs, iMER consider the three properties to categorize the attributes i.e., type of operation, subject, and object.

*Dependency: nsubject(A,B), nsubjpass(A,B), dobj(A,B), iobj(A,B), pobj(A,B), nmod:to(A,B) and mark(A,B)*

Part 'A' of these TDs contains verb (operation) and B is a noun (entity or attribute). If the verb is considered as an input operation (e.g., enter, insert, input, select, click, choose), then the attribute is input (*TDR 27 in Appendix*).

*Example*: "User enters ID and password to login."
→ *nsubj(enter-2, user-1)*

This example contains a verb "*enter*", indicating that attributes "*ID and password*" are inputs of the login function.

If verbs represent the output, e.g., output, display, retrieve, then the attributes are the outputs of an operation (*TDR 28 in Appendix*).

*Example:* "The system displays the sales report containing date, product ID, product name, price and total amount for the requested date."

The TD extracted from the above sentence is *nsubj(display-2, System-1),* verb "d*isplay*" indicates that data in this example is output. For some verbs, it is difficult to decide (e.g., get, send, prepare etc.). In this situation, iMER also considers the nouns in subject and object (*TDR 29 in Appendix*).

*Example*: "System gets the date for searching."
→ *nsubj(gets-2, System-1)*
"Customer gets the code to verify."
→ *nsubj(gets-2, Customer-1)*

The above TDs have the same operation "gets" and we have to look further into the subjective part i.e., "System" in the first sentence and "Customer" in the second, are categorized as input and output respectively.

*Dependencies: nmod:by(A,B), nmod:agent(A,B), nmod:with(A,B)*

Based on the verbs in A of these TDs, the attributes can be categorized. If the verb indicates an input operation, then the attributes in previous TDs are inputs (*TDR 30 in Appendix*).

*Example:* "Name and address are entered by the customer"
→ *nmod:by(entered-4, customer-7)*

If verb in A represents an output operation, then the attribute in previous TDs is the output data (TDR 31 in Appendix).

*Example*: "Transaction no will be displayed by the system"
→ *nmod:agent ( displayed-5 , system-8 )*

### 3.5. Step 5: Business Process Model Extraction

BP model is to illustrate the detailed operations of required functionality. It depicts the relationships between external (user and other systems that interact with this system) and internal (system) operations. It is beneficial if the operation presented with its input and output. As most of the operations extracted from the requirement are described in single word, so the additional information such as data consumed can help to understand the required functionality.

This iteration of iMER is to generate all possible flows for the BP model (depending on the requirement format). In the model, there are five types of flows. Two of them, <<External Action>> and <<System Action>> are common in all requirements while rest of them depend on the requirements format.

a. <<External Action>>: This flow contains the actions performed by the external actors (users or other systems that interact with the proposed system).
b. <<Alternate External Action>>: These actions can be extracted from the alternate section of UCS. This flow is parallel to the main flow. It represents the alternate or optional choices of execution. For instance, user has a choice of entering the data (in main flow) or cancel the operation (in alternate). Therefore, "Enter data" and "Cancel" are the parallel external operations.
c. <<System Actions>>: This path contains the expected action performed by the system (internal operations).
d. <<Alternate System Actions>>: Commonly, these operations appear in the alternate section of UCS. This is the alternative or optional part of system operations, for example, in case of conditional statement, parallel paths for true and false conditions.
e. << Exception>>: It is another parallel path to the system operations. If something wrong happened in system or that something the system operation could not handle, an exception will be thrown. This path contains only the error handling / error messages / exceptions.

Each system operation is represented with its input/output and interaction with the data entity. The following figure depicts the system operations w.r.t the flow paths.

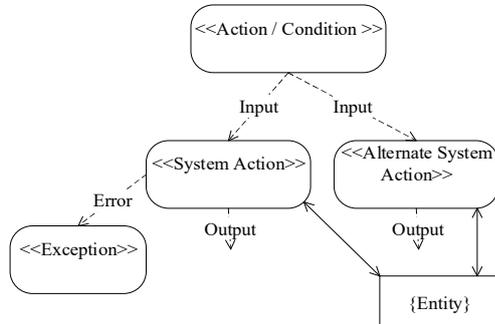

**Fig. 4** Representation of System Operation in BP model

### 3.5.1. User and System operation extracted in BP model

The following operations are extracted from the requirements to complete all the internal and external flow paths.

#### 3.5.1.1. Basic operations

Basic operations are inputting data, processing or generating output

*Dependencies: nsubj(A, B), nmod:by(A, B)*

In these TDs, A is a verb (operation/action) and B is noun. If verb of a sentence is an input action (e.g., enter, insert, input, select, click or choose) and subject is an external actor, then it is a <<External Action>>. The attributes attached to this operation will be the input of next system operation. If sentence is in the alternate section of UCS, then the action will be added to the <<Alternate Action>> path (*TDR 32 in Appendix A*).

*Example:* "A customer clicks on the product_no to view the details."
  root (ROOT-0 , clicks-3), det(customer-2 , A-1), nsubj(clicks-3 , customer-2), case(product_no-6 , on-4), det(product_no-6 , the-5), nmod:on(clicks-3 , product_no-6), mark ( view-8 , to-7 ), acl(product_no-6 , view-8), det(details-10 , the-9), dobj(view-8 , details-10)

In this example, the TD *nsubj(clicks-2 , customer-1)* is indicating that this action belongs to the <<External Action>> path along with data '*product_no*'.

If the verbs from these TDs is an output operation (e.g., display, output, retrieve, show, view, print) or a data processing operation (e.g., calculate, process, update, delete, search, modify, edit, calculate, remove) and the subject is system, then it is an internal operation and will be added in the <<System Actions>> path (*TDR 32 in Appendix A*).
*Example:* "System displays the product details."
  TDs generated are:
  root(ROOT-0, display-2), nsubj ( display-2, System-1 )
  det(details-5, the-3 ), compound ( details-5, product-4 )
  dobj ( display-2 , details-5 )

Operation "*display details (product_no)*" will be added to the system operation path. Variable '*product_no*' from the previous external operation is the input of this system operation.

If A contains any of these actions (receive, accept, get, obtain, acquire) and the B (subject) is system, then it will be taken as an external operation or internal in case of a passive voice (*TDR 33 in Appendix A*).
*Example*: "System receives the data to process."
root ( ROOT-0, receive-2 ), nsubj(receive-2, System-1 )
det(data-4, the-3 ), dobj(receive-2, data-4), case(process-6 , to-5 ), nmod:to ( receive-2 , process-6 )
Verb "*receive*" in *nsubj(receive-2, System-1 )* suggests it is an input to the system.

#### 3.5.1.2. Errors (Exceptions)

If a statement contains any expected error message, then it will be added to the exceptions path. In UCS, exception section contains all the error messages. Also, the following TDs can help to extract the exceptions.

*Dependencies: xcomp (A, B), dobj (A, B), neg(A,B)*

If A or B contains any of the keyword words (error, fail, wrong, invalid, incorrect, not etc.), then it might be an error message (*TDR 34 in Appendix A*).
*Example:* "System failed to display products."
TD *xcomp ( failed-2 , display-4 )* generates '*failed display*' along with the entity '*Product*'

#### 3.5.1.3. Conditional statements

Conditional statements are the internal statements. These are the conditional checks performed by the system.

*Dependencies: advcl:if(A, B), mark(A, if), advmod(A, then), advmod(A, else), dobj(A,B)*

These TDs can help to extract the conditional statements. The advcl:if(A,B) contains verbs in both A & B parts. B part of this TD will be the conditional checking and if the condition is true, then operation in A will be executed. Verbs in the '*nsubj*' or '*dobj*' will also be required to complete the true part of condition (*TDR 35 in Appendix A*).
*Example:* "If system founds the reservation then it will display the customer name."
TDs generated are: mark(found-3, if-1), nsubj(found-3, syste-2), advcl:if(display-9, found-3), dobj(found-3, reservation-5), advmod(found-3, then-6), dobj(display-9, name-12)
From these TDs, we got an internal operation "*if found then display*" with an entity '*reservation*' and attribute '*name*'.

*Dependencies: nsubj(validate, system)*

This TD is also used to find the conditional checking (*TDR36 in Appendix A*).
Example: "System validates the username and password."
→ *nsubj (validates-2 , System-1 )* contains "*validates*" conditional check with attributes '*username*' and '*password*'.

### 3.5.1.4. Iterations

It is to know that which operations are to be repeated based on a condition.

*Dependencies: nsubj(A,B), xcomp(A,B), nummod(A,B), dobj(A,B)*

If *nsubj* or *xcomp* TDs contains any of these verbs (continue, restart, go, repeat, move, jump) then by considering the nmod or dobj TDs, iMER can indentify the set of statements to be repeated (*TDR 37 in Appendix A*).
*Example:* 4. "Use case continue at step 1."
*nsubj (continue-3, step-5), nummod(step-5, 1-6)* are indicating a jump to statement #1 from the current statement.

## 4. Case Studies

In this section, we are presenting several case studies to illustrate the process of iMER and possible variations to these steps based on the type of the requirement document. The requirements are taken from the published literatures to compare the results.

### 4.1. Case Study # 1

In this case study, general requirements of "Online Order" are processed to generate the artefacts.

*"The system shall display the shopping cart during online purchase. The system shall allow user to add products in the shopping cart. The system shall display the products from the shopping cart. The system shall display different shipping options. The system shall enable user to select the payment method, billing address and shipping method of the order during payment process. The system shall display the shipping charges. The system shall display tentative duration for shipping. The system shall allow user to enter the order ID for tracking. The system shall display the dispatch date, time and current status about the order. The system shall allow user to confirm the purchase. The system shall enable user to enter the payment method, billing address, shipping address and shipping method during payment process. The system shall calculate tax for the order. The system shall display tax information for the order. The system shall update the payment."*

In iMER, the Stanford CoreNLP 3.8 API is used for the preprocessing (tokenization, lemmatizing, POS-tagging and TDs generation) of each sentence. iMER used TDRs (listed in the annexure A.1) for the extracting of entities and attributes in its first iteration. Table 3 contains the entities extracted from the above-mentioned requirements. First column contains the entities while the second has the frequency of each entity (number of times an entity extracted from the TDs).

**Table 3** Entities extracted from the requiremnts

| Entities | Frequency (w.r.t TDs) |
|---|---|
| shopping cart | 7 |
| customer | 10 |
| product | 3 |
| payment | 3 |
| shipping | 6 |
| order | 5 |
| tracking | 1 |

Table 4 contains the attributes of the respective entities extracted.

**Table 4** Attributes of the entities extracted from case study # 1

| Entities | Attributes |
|---|---|
| payment | method |
| shipping | shipping method, shipping charges, tentative duration |
| order | id, tax |
| tracking | date, time, current status |

After the extraction of entities and attributes, TDRs (*listed in the Appendix A.2*) are applied to extract the relationships. In iMER, the TDRs (*listed in Appendix A.3*) are used to find the cardinalities in its third iteration. In the process, only those TDs that have the entities are processed and this improved the performance of iMER. Table 5 contains the identified relationships between entities.

**Table 5** Entities Relationships generated from the case study # 1

| |
|---|
| product▷* (add) customer▷1 |
| shopping cart▷1 (add) customer▷1 |
| shopping cart▷1 (has) product▷* |
| shipping▷1 (has) order▷1 |
| shipping▷1 (select) customer▷1 |
| payment▷1 (select) customer▷1 |
| payment▷1 (has) shipping▷1 |
| tracking▷1 (enter) customer▷1 |
| tracking▷1 (has) order▷* |
| order▷1 (confirm) customer▷1 |

ER model depicted in Fig. 5 is the graphical representation of the compoenets, extracted from the requirements presented in our first case study by applying the TDRs.

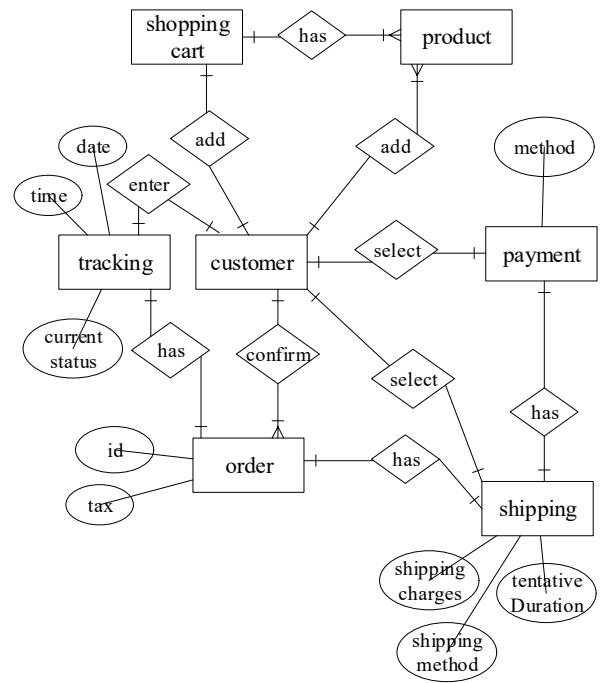

**Fig. 5.** ER model extracted from requirements in case study # 1

Table 6 contains the inputs and output of '*online purchase*' functionality.

**Table 6** Catagorized Attributes

| Input | Output |
|---|---|
| payment method, shipping method, order id | shipping charges, duration, date, time, current status, tax |

Next iteration of iMER is to extract the BP model. The model illustrates the complete flow of data along with operations. The BP model in Fig.6 is extracted from the requirements of first case study. In this BP model, rectangles with rounded corners are to represent the operations. Actions performed by the user have the <<External Action>> stereotype, while system actions are denoted by <<System Action>>. Open-ended rectangles are to represent the data entities and solid arrows are to show the flow of data while dashed arrows are for the control flow.

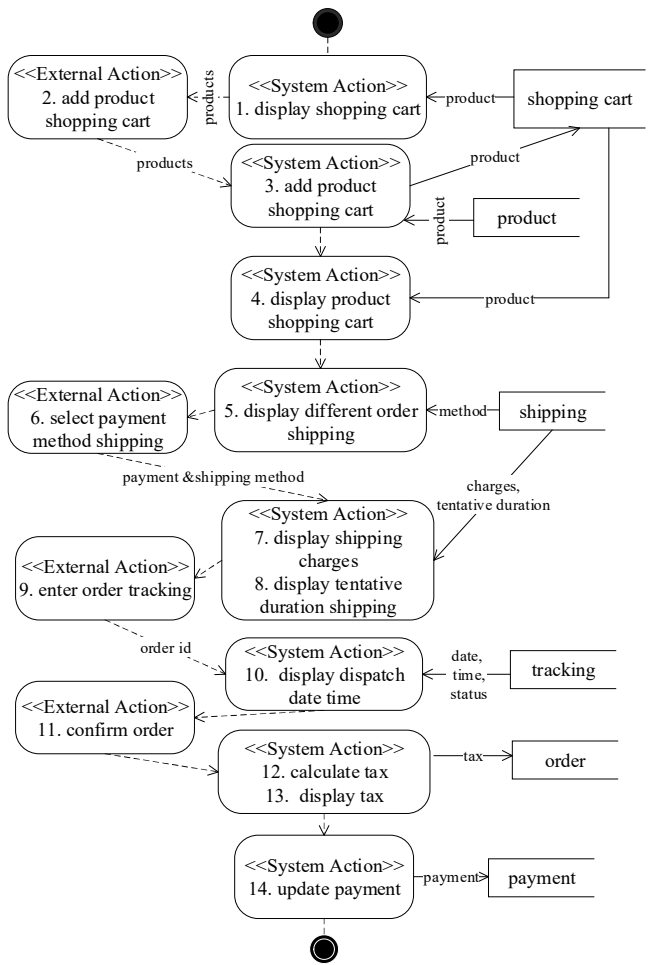

**Fig. 6.** BP model generated from the requirements in case study # 1

## 4.2. Case Study # 2

This case study is to process the user stories for the extraction of artefacts.

*As a Visitor, I can create a new account.*
*As a Visitor, I can log in.*
*As a Visitor, I can change my account password.*
*As a Visitor, I am able to search for an event.*
*As a Visitor, I want to filter on event type, so that I can only see events of the type I want.*
*As a Visitor, I want to choose an event so that I can book a ticket for that event.*
*As a Visitor, I want to see the ticket price.*
*As a Visitor, I want to choose a type of ticket.*
*As a Visitor, I am able to purchase multiple tickets.*
*As a Visitor, I want to provide my personal details to purchase a ticket.*
*As a Visitor, I want to choose a payment method so that I can buy a ticket.*
*As a Visitor, I want to receive a purchased ticket."*

After the pre-processing, TDRs applied to extract the entities (listed in Table 7) and attributes (in Table 8) in its first iteration. For a comparison, these user stories also processed using the tool proposed in [2] by setting the frequency threshold. Lucassen et al. [2] generated the entities (*Visitor, Ticket, Event, Account, Type, I, Detail, Method, Price* and *Password*) using the Visual Narrator tool. Comparing to [2], we got the better results, presented in the following tables.

**Table 7** Entities extracted from the user stories in case study # 2

| Entities | Frequency (w.r.t TDs) |
|---|---|
| visitor | 12 |
| event | 7 |
| ticket | 8 |
| payment | 1 |

**Table 8** Attributes of the entities generated from the case study # 2

| Entities | Attributes |
|---|---|
| visitor | password |
| event | type |
| ticket | price |
| payment | method |

Next iterations are to extract the relationships and cardinalities. The relationships extracted by analyzing the transitive verbs. The relationships extracted from the user stories are presented in Table 9.

**Table 9** Entities Relationships generated from the case study # 2

| |
|---|
| visitor>1 (search) event>1 |
| visitor>1 (see) event>* |
| visitor>1 (has) event>* |
| visitor>1 (choose) event>1 |
| ticket>1 (has) event>1 |
| visitor>1 (see) ticket>1 |
| visitor>1 (purchase) ticket>* |
| visitor>1 (provide) ticket>1 |
| visitor>1 (choose) payment>1 |

visitor>1 (choose) ticket>1
ticket>1 (has) payment>1
visitor>1 (receive) ticket>1

In ER model, same pair of data entities might have the multiple relationships. The relationships between a pair of data entities are collectively represented in one diamond. ER model extracted from the case study # 2 is illustrated in Fig. 7.

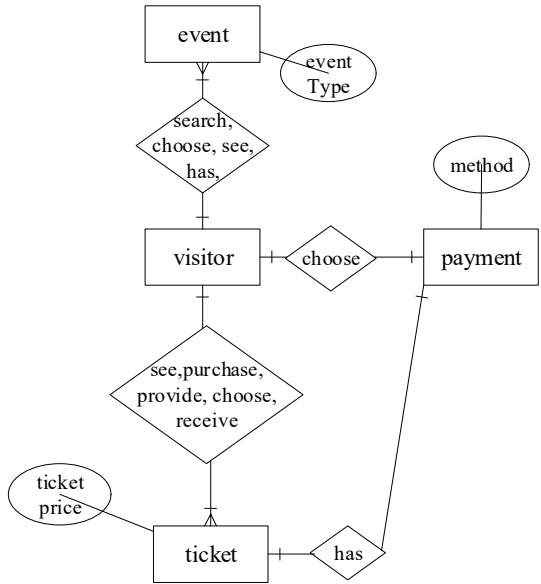

**Fig. 7.** ER model generated from the user stories mentioned in case study # 2

Operations in the user stories are the actions required by the users. Fig. 9 (in annexure C) contains the operations extracted from these user stories.

**4.3. Cast Study # 3**

This case study contains a UCS of "Capture witness Report" from Car Crisis management system from [33].

*Scope: Car Crash Crisis Management System.*
*Primary Actor: Coordinator.*
*Secondary Actor: PhoneCompany, Surveillance System*
*Intention: The Coordinator intends to create a crisis record based on the information obtained*
*from witness.*
*Main Success Scenario.*
*1. Coordinator provides witness details (first name, last name, phone number, and address) to System as reported by the witness.*
*2. Coordinator informs System of crisis location and type as reported by the witness.*
*3. System contacts PhoneCompany to verify witness information.*
*4. PhoneCompany provides the witness address and phone information to System.*
*5. System validates information received from the PhoneCompany.*
*6. System provides Coordinator with a crisis-focused checklist.*
*7. System starts displaying Surveillance video feed for Coordinator.*
*8. Coordinator provides crisis information (crisis details, crisis time) to System as reported by the witness.*
*9. System assigns an initial emergency level to the crisis and sets the crisis status to active. Use case ends in success.*

*Alternate.*
*1. The call is disconnected. The base use case terminates.*
*2. The call is disconnected. The base use case terminates.*
*5. PhoneCompany information does not match information received from Witness.*
*8. The call is disconnected.*

Firstly, iMER aligned sentences from the alternate section with the relevant sentences in the main flow section (sentence sequencing). For instance, the first sentence in the alternate flow, i.e., '1a' is inserted after the sentence # 1 of the main flow to complete the branching processes. After the sequencing of sentences and preprocessing, iMER applied TDRs to extract the entities (in Table 10) and attributes (in Table 11).

**Table 10** Entities extracted from the UCS in case study # 3

| Entities | Frequency (w.r.t TDs) |
|---|---|
| coordinator | 5 |
| Witness | 8 |
| Crisis | 9 |
| Phonecompany | 1 |
| Checklist | 1 |
| surveillance | 1 |
| Emergency | 2 |

**Table 11** Entitie's attributes extracted from the UCS in case study# 3

| Entities | Attributes |
|---|---|
| witness | first name, last name, phone number, witness address |
| Crisis | Crisis location, type, time, crisis status |
| Emergency | emergency level |

After applying the TDRs (in appendices A.2 & A.3), relationships and cardinalities are generated in the second and third iterations of iMER respectively. Results are presented in the Table 12.

**Table 12** Entities Relationships generated from the case study # 3

coordinator>1 (reported) witness>1
coordinator>1 (provides) witness>1
coordinator>1 (informs) crisis>1
phonecompany>1 (match) witness>1
coordinator>1 (provides) phonecompany>1
surveillance>1 (starts) coordinator>1
coordinator>1 (provides) checklist>1
emergency>1 (sets) crisis>1

Using the extracted elements of artefact from the case study # 3, an ER model is illustrated in Fig. 8.

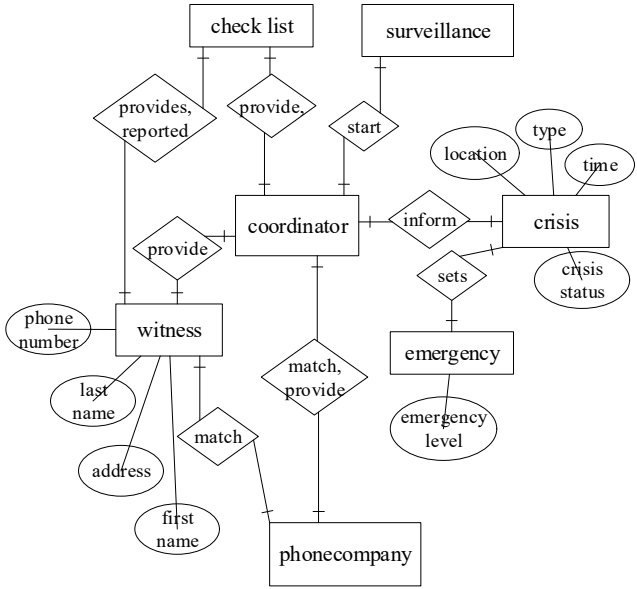

**Fig. 8.** ER model genrated from UCS mentioned in case study 3

The BP model extracted from the "*Capture Witness Report*" UCS is illustrated in Fig. 10 (in annexure C). The process model shows the complete flow of data with operations. As discussed in Section 3.5, BP extracted from the UCS could have some extra flows parallel to the user action and system actions.

## 5. Evaluation

In this section, our objective is to verify that how efficient this technique is for extracting the conceptual models from the requirement documents.

### 5.1. Data set

The proposed technique is implemented and tested on various types of requirement specifications including general requirements, user stories and UCSs. General requirements stated in different templates are processed, that include but not limited to;

- *IEEE Template*: It is to present the requirements in a defined structure of the document.
- *General Template*: In general format, mostly requirements are stated in a sequential way.
- *Descriptive Template*: is to describe the functionality in a conversational manner.
- *Paragraph Template*: In this format modules of the proposed system are explained in different paragraphs.

A sample data set of these requirements formats are presented in annexure B.1.

As stated earlier in Section 1 that UCSs could have different or extended templates. These templates include different sections and synonyms section headings depending on the organizational practices. Some of the sample templates of UCs are presented in annexure B.2.

These requirements are collected in various sizes from the different sources which includes;

- Professional institutions, because these are related to the real-world problems and structured documents.
- Final projects of the computer science graduate students, because most of them do not follow a specific format, these are the most unpredictable structures of requirements.
- Requirements processed by the other researchers to validate their approaches. These documents helped us to compare our technique with the exiting.

### 5.2. Evaluation Matric

We used the same performance metrics as used by the researchers for exiting techniques, i.e., recall and precision. To differentiate the projected values and actual values, percentile splitting of three classifications, i.e., True Positive (TP), False Positive (FP) and False Negative (FN) are presented. Derived metrics for performance measurement are;

*Recall (RCL)*: to measure the correctly extracted elements by the proposed solution [31].

$$RCL\% = \frac{TP}{TP + FN} * 100$$

*Precision (PRC)*: to measure the ratio between correct element generated by the approach and incorrect elements [31].

$$PRC\% = \frac{TP}{TP + FP} * 100$$

$F_1$ *Score*: is the harmonic mean of precision and recall [31].

$$F1\% = 2 * \frac{Precision * Recall}{Precision + Recall} * 100 = \frac{2 * TP}{2TP + FP + FN} * 100$$

### 5.3. Results and Discussion

The main objective is to verify that "*Can iMER extract the required artefacts from different structures of sentences?*" Therefore, these requirements are processed in three different ways, i.e., a) without changing, b) shuffling the sentences and c) merging the requirements from different documents of same domain. The cases in b) and c) represent the poor structured documents and more complex sentence structures.

All documents are inputted in plain text files (.txt file format). Information extraction system is normally evaluated by comparing its output with the manual results [6]. Therefore, artefacts generated by the tool are evaluated manually.

#### 5.3.1. Validation results without changing the requirements

Tables 13-18 present the results of ER models extraction from the general requirements documented in different

formats. In these experiments, we processed the requirements without changing them.

**Table 13** Accuracy for e-Store Requirements (IEEE830 format)

|  | TP % | FP % | FN % | RCL % | PRC% | $F_1$% |
|---|---|---|---|---|---|---|
| Entities | 87.8 | 4.9 | 7.3 | 92 | 95 | 94 |
| Attributes | 88.7 | 4.8 | 6.5 | 93 | 95 | 94 |
| Relationships | 84.4 | 9.4 | 6.2 | 93 | 90 | 92 |
| Cardinalities | 89.3 | 3.6 | 7.1 | 93 | 96 | 94 |

**Table 14** Table 14 Accuracy for e-Store Requirements (general format)

|  | TP % | FP % | FN % | RCL % | PRC% | F1% |
|---|---|---|---|---|---|---|
| Entities | 87.5 | 3.1 | 9.4 | 90 | 97 | 93 |
| Attributes | 84.9 | 9.4 | 5.7 | 94 | 90 | 92 |
| Relationships | 86.4 | 9.1 | 4.5 | 95 | 90 | 93 |
| Cardinalities | 84.2 | 10.5 | 5.3 | 94 | 89 | 91 |

**Table 15** Accuracy for e-Store Requirements (Paragraph format)

|  | TP % | FP % | FN % | RCL % | PRC% | $F_1$% |
|---|---|---|---|---|---|---|
| Entities | 89.3 | 3.6 | 7.1 | 93 | 96 | 94 |
| Attributes | 87.2 | 5.1 | 7.7 | 92 | 94 | 93 |
| Relationships | 83.3 | 5.6 | 11.1 | 88 | 94 | 91 |
| Cardinalities | 81.3 | 12.5 | 6.2 | 93 | 87 | 90 |

**Table 16** Accuracy for e-Store Requirements (descriptive format)

|  | TP % | FP % | FN % | RCL % | PRC% | $F_1$% |
|---|---|---|---|---|---|---|
| Entities | 87.0 | 8.7 | 4.3 | 95 | 91 | 93 |
| Attributes | 88.1 | 6.8 | 5.1 | 95 | 93 | 94 |
| Relationships | 83.3 | 4.2 | 12.5 | 87 | 95 | 91 |
| Cardinalities | 84.2 | 5.3 | 10.5 | 89 | 94 | 91 |

**Table 17** Accuracy for Air Ticket Reservation (descriptive format)

|  | TP % | FP % | FN % | RCL % | PRC% | $F_1$% |
|---|---|---|---|---|---|---|
| Entities | 85.0 | 10.0 | 5.0 | 94 | 89 | 92 |
| Attributes | 89.3 | 7.1 | 3.6 | 96 | 93 | 94 |
| Relationships | 85.0 | 5.0 | 10.0 | 89 | 94 | 92 |
| Cardinalities | 85.7 | 4.8 | 9.5 | 90 | 95 | 92 |

**Table 18** Accuracy for Air Ticket Reservation (paragraph format)

|  | TP % | FP % | FN % | RCL % | PRC% | $F_1$% |
|---|---|---|---|---|---|---|
| Entities | 89.3 | 7.1 | 3.6 | 96 | 93 | 94 |
| Attributes | 86.0 | 9.3 | 4.7 | 95 | 90 | 93 |
| Relationships | 85.3 | 5.9 | 8.8 | 91 | 94 | 92 |
| Cardinalities | 84.4 | 6.3 | 9.3 | 90 | 93 | 92 |

Tables 19-22 are to present the validation results of ER models extraction from UCSs with different templates.

**Table 19** Accuracy for e-store System (UCS template 1)

|  | TP % | FP % | FN % | RCL % | PRC% | $F_1$% |
|---|---|---|---|---|---|---|
| Entities | 87.2 | 5.1 | 7.7 | 92 | 94 | 93 |
| Attributes | 84.0 | 10.0 | 6.0 | 93 | 89 | 91 |
| Relationships | 87.7 | 5.3 | 7.0 | 93 | 94 | 93 |
| Cardinalities | 94.0 | 2.0 | 4.0 | 96 | 98 | 97 |

**Table 20** Accuracy for e-store System (UCS template 2)

|  | TP % | FP % | FN % | RCL % | PRC% | $F_1$% |
|---|---|---|---|---|---|---|
| Entities | 83.9 | 9.7 | 6.4 | 93 | 90 | 91 |
| Attributes | 88.3 | 7.0 | 4.7 | 95 | 93 | 94 |
| Relationships | 81.4 | 11.6 | 7.0 | 92 | 88 | 90 |
| Cardinalities | 83.8 | 10.8 | 5.4 | 94 | 89 | 91 |

**Table 21** Accuracy for Ticket Reservation System (UCS template 1)

|  | TP % | FP % | FN % | RCL % | PRC% | $F_1$% |
|---|---|---|---|---|---|---|
| Entities | 88.0 | 8.0 | 4.0 | 96 | 92 | 94 |
| Attributes | 86.1 | 8.3 | 5.6 | 94 | 91 | 93 |
| Relationships | 83.3 | 9.5 | 7.2 | 92 | 90 | 91 |
| Cardinalities | 79.5 | 11.4 | 9.1 | 90 | 88 | 89 |

**Table 22** Accuracy for Ticket Reservation System (UCS template 2)

|  | TP % | FP % | FN % | RCL % | PRC% | $F_1$% |
|---|---|---|---|---|---|---|
| Entities | 85.3 | 8.8 | 5.9 | 94 | 91 | 92 |
| Attributes | 83.3 | 9.5 | 7.2 | 92 | 90 | 91 |
| Relationships | 84.8 | 9.1 | 6.1 | 93 | 90 | 92 |
| Cardinalities | 89.7 | 6.9 | 3.4 | 96 | 93 | 95 |

Tables 23 & 24 are to presents the results of user stories. A sample of user stories is presented in the Section 4.2 as a case study.

**Table 23** Accuracy for the user stories of e-Store System

|  | TP % | FP % | FN % | RCL % | PRC% | $F_1$% |
|---|---|---|---|---|---|---|
| Entities | 83.3 | 10.0 | 6.7 | 93 | 89 | 91 |
| Attributes | 80.8 | 11.5 | 7.7 | 91 | 88 | 89 |
| Relationships | 87.2 | 7.7 | 5.1 | 94 | 92 | 93 |
| Cardinalities | 89.5 | 7.9 | 2.6 | 97 | 92 | 94 |

**Table 24** Accuracy for the Web Gold user stories

|  | TP % | FP % | FN % | RCL % | PRC% | $F_1$% |
|---|---|---|---|---|---|---|
| Entities | 86.8 | 7.9 | 5.3 | 94 | 92 | 93 |
| Attributes | 88.5 | 7.7 | 3.8 | 96 | 92 | 94 |
| Relationships | 86.7 | 8.9 | 4.4 | 95 | 91 | 93 |
| Cardinalities | 88.1 | 7.1 | 4.8 | 95 | 93 | 94 |

Following tables contains the validation results for the extraction and categorization of the BP model's components. We validated the identified user and system operations (operation categorization) and data for these operations. Values of FP for the operations are based on the valid separation of user and system operations. While for the data categorizations, it is based on the separation of input and output. The results demonstrate a noticeable improvement.

**Table 25** Accuracy for e-store System (UCS template 1)

| Categorizations | TP % | FP % | FN % | RCL % | PRC% | $F_1$% |
|---|---|---|---|---|---|---|
| Operations | 88.6 | 6.8 | 4.6 | 95 | 93 | 94 |
| Data | 86.1 | 8.3 | 5.6 | 94 | 91 | 93 |

**Table 26** Accuracy for e-store System (UCS template 2)

| Categorizations | TP % | FP % | FN % | RCL % | PRC% | $F_1$% |
|---|---|---|---|---|---|---|
| Operations | 90.9 | 5.5 | 3.6 | 96 | 94 | 95 |
| Data | 85.3 | 8.8 | 5.9 | 94 | 91 | 92 |

**Table 27** Accuracy for Ticket Reservation System (UCS template 1)

| Categorizations | TP % | FP % | FN % | RCL % | PRC% | $F_1$% |
|---|---|---|---|---|---|---|
| Operations | 82.4 | 9.8 | 7.8 | 91 | 89 | 90 |
| Data | 79.5 | 12.8 | 7.7 | 91 | 86 | 89 |

**Table 28** Accuracy for Ticket Reservation System (UCS template 2)

| Categorizations | TP % | FP % | FN % | RCL % | PRC% | $F_1$% |
|---|---|---|---|---|---|---|
| Operations | 85.7 | 8.6 | 5.7 | 92 | 88 | 90 |
| Data | 81.5 | 11.1 | 7.4 | 92 | 88 | 90 |

### 5.3.2. Validation results after shuffling the sentences of requirements

We processed the requirement documents containing the sentence structures ranging from simple to the complex level. Sentences are shuffled for validation purpose. We used automated sentence shuffling software. The output of this evaluation is compared with the results produced in the first evaluation (i.e., without changing the documents). These experiments demonstrated that changing the sequence of sentences would not affect the results of ER model extraction significantly because each sentence in the requirements documents is processed separately. Results after shuffling the sentences are almost similar to the original cases in Section 5.1. The relationship extraction is based on the flow of data, as the flow of data is disturbed by changing the sequence of sentences, so we see that the accuracy of dropped as expected. User activities and system operations are separated but the process flow is disturbed due to shuffled requirements.

### 5.3.3. Validation results after merging the requirements

As our technique is not restricted to a specific format of requirements. Therefore, in this evaluation step, we merged the requirement documents from the same domain, to test the robustness of technique. Documents collected from various sources contain different level of details and the structures of the requirements. Naturally, it is more difficult to accurately extract the artefacts. Like previous, output of this step is also compared with the results from first evaluation. Results showed negligible differences as compared to processing of the requirement documents separately. Tables listed below (29 to 31) contain the validation of results of ER models extraction from the merged requirements.

**Table 29** Accuracy for e-store merged requirements (IEEE and general format)

|               | TP % | FP % | FN % | RCL % | PRC% | $F_1$% |
|---------------|------|------|------|-------|------|--------|
| Entities      | 81.6 | 10.2 | 8.2  | 91    | 89   | 90     |
| Attributes    | 83.8 | 7.4  | 8.8  | 90    | 92   | 91     |
| Relationships | 80.0 | 8.9  | 11.1 | 88    | 90   | 89     |
| Cardinalities | 82.9 | 9.8  | 7.3  | 92    | 89   | 91     |

**Table 30** Accuracy for e-store merged requirements (Paragraph and Descriptive format)

|               | TP % | FP % | FN % | RCL % | PRC% | $F_1$% |
|---------------|------|------|------|-------|------|--------|
| Entities      | 83.6 | 9.1  | 7.3  | 92    | 90   | 91     |
| Attributes    | 84.5 | 8.5  | 7.0  | 92    | 91   | 92     |
| Relationships | 79.4 | 11.8 | 8.8  | 90    | 87   | 89     |
| Cardinalities | 77.4 | 12.9 | 9.7  | 89    | 86   | 87     |

**Table 31** Accuracy for merged Air Ticket requirements (Paragraph and Descriptive format)

|               | TP % | FP % | FN % | RCL % | PRC% | $F_1$% |
|---------------|------|------|------|-------|------|--------|
| Entities      | 84.8 | 9.1  | 6.1  | 93    | 90   | 92     |
| Attributes    | 81.3 | 10.4 | 8.3  | 91    | 89   | 90     |
| Relationships | 80   | 11.4 | 8.6  | 90    | 89   | 92     |
| Cardinalities | 85.0 | 10.0 | 5.0  | 94    | 89   | 92     |

Tables 32 & 33 contains the validation of results for merged UCSs. Although the sequence of flow-graph is affected but the entities and attributes identified are merely changed.

**Table 32** Accuracy for merged e-store UCSs (templates 1 & 2)

|               | TP % | FP % | FN % | RCL % | PRC% | $F_1$% |
|---------------|------|------|------|-------|------|--------|
| Entities      | 82.7 | 7.7  | 9.6  | 90    | 91   | 91     |
| Attributes    | 79.7 | 10.9 | 9.4  | 89    | 88   | 89     |
| Relationships | 81.5 | 9.9  | 8.6  | 90    | 89   | 90     |
| Cardinalities | 78.5 | 11.4 | 10.1 | 89    | 87   | 88     |

**Table 33** Accuracy for merged Air Ticket System UCSs (templates 1 & 2)

|               | TP % | FP % | FN % | RCL % | PRC% | $F_1$% |
|---------------|------|------|------|-------|------|--------|
| Entities      | 77.3 | 13.6 | 9.1  | 89    | 85   | 87     |
| Attributes    | 78.0 | 12.0 | 10.0 | 89    | 87   | 88     |
| Relationships | 80.0 | 11.4 | 8.6  | 90    | 88   | 89     |
| Cardinalities | 81.0 | 8.6  | 10.3 | 89    | 90   | 90     |

Table 34 is to presents the validation results of merged user stories. In this evaluation, we merged the user stories collected from two different sources, i.e., e-store system and web gold. The proposed technique efficiently extracted all the artefacts.

**Table 34** Accuracy for the merged user stories

|               | TP % | FP % | FN % | RCL % | PRC% | $F_1$% |
|---------------|------|------|------|-------|------|--------|
| Entities      | 83.9 | 7.2  | 8.9  | 90    | 92   | 91     |
| Attributes    | 83.3 | 10.0 | 6.7  | 93    | 89   | 91     |
| Relationships | 83.8 | 8.8  | 7.4  | 92    | 91   | 91     |
| Cardinalities | 84.7 | 8.3  | 6.9  | 92    | 91   | 92     |

Tables 35 & 36 contains the results of BP models extraction after merging the requirements. iMER efficiently extracted the components of BP model but the sequencing of operation affected and, understandably, resulted in some branching problems.

**Table 35** Accuracy for merged e-store UCSs (templates 1 & 2)

|                 | TP % | FP % | FN % | RCL % | PRC% | $F_1$% |
|-----------------|------|------|------|-------|------|--------|
| Categorizations |      |      |      |       |      |        |
| Operations      | 80.2 | 11.5 | 8.3  | 91    | 88   | 89     |
| Data            | 80.9 | 8.8  | 10.3 | 89    | 90   | 89     |

**Table 36** Accuracy for merged Air Ticket System UCSs (templates 1 & 2)

|                 | TP % | FP % | FN % | RCL % | PRC% | $F_1$% |
|-----------------|------|------|------|-------|------|--------|
| Categorizations |      |      |      |       |      |        |
| Operations      | 78.7 | 12.0 | 9.3  | 89    | 87   | 88     |
| Data            | 79.5 | 11.4 | 9.1  | 90    | 88   | 89     |

### 5.4. Comparative study

This section contains the comparison of results with the existing tools. For a comparison, we processed the same set of requirements as in [2]. Table 37 is the comparison with the proposed approach of Lucassen et al. [2] by processing the same user stories, i.e., ArchivesSpace. Although the percentile of false negative in relationships extraction is slightly higher due to the mismatched verbs, however, the overall results of our tool are showing improvements for the extraction of ER model from user stories.

**Table 37** Comparison of Performance Metrics based on User Stories of Archivespace

|  | Entities | | | Relationships | | |
|---|---|---|---|---|---|---|
|  | TP % | FP % | FN % | TP % | FP % | FN % |
| [2] | 89.5 | 3.9 | 6.5 | 74.4 | 10.9 | 14.7 |
| **iMER** | 92.3 | 3.4 | 4.3 | 79.0 | 5.5 | 15.5 |

Table 38 contains the comparison with the results of existing tools, presented in [30] by processing the same requirements i.e., library management system. It showed that the results of our tool are improved as compared to the exiting tools.

**Table 38** Comparison of Performance Metrics based on the Requirements of Library System

|  | [19] | [21] | [25] | [28] | [30] | [37] | **iMER** |
|---|---|---|---|---|---|---|---|
| RCL% | 62 | 73 | 72 | 56 | 89 | 52 | 95 |
| PRC% | 70 | 66 | 65 | 65 | 92.3 | 75 | 93 |

In all these evaluations, our tool iMER shows better results. However after shuffling the requirements, results related to the relationship extraction fluctuated that are understandable. The improvements are due to the number of reasons discussed below.

### 5.5. Discussion

In this section, we elborate on the reasons of better results generated by our approach.

#### 5.5.1. Detailed Analysis

As discussed earlier (Section 1) that in this approach, we processed various formats of requirement documents with different sentence sturctures ranging from simple to complex level. Therefore, requirements are analyzed at a greater level of detail. We processed more TDs as compared to the exiting techniques. Table 39 is to show TDs processed in iMER vs. processed in other approaches. For instance, passive nominal subject (*nsubjpass*) is only processed by [6] & [12]. Similarly, indirect object (*iobj*) is processed only in one approach [12], object of preposition (*pobj*) is considered in [2] & [12]. The nominal modifier (*nmod*) relation holds between the noun / predicate modified by the prepositional complement. From these TDs, [12] & [15] considered some prepositional relation. While, in [6] & [25], only one nominal modifier is processed, but no one considered *nmod:from, nmod:as,* nmod:by, & nmod:with, because simple or specific stuctures of the senetnces are assumed. Same in the case with other TDs, conjunct (*conj:and & conj:or*), determiner (*det*), and adverbial clause modifier (*advcl:if*); none of them is processed in the exiting technqiues. Some relations (TDs) presented in exiting techniques were generated by using the older version of Stanford CoreNLP API but, these are removed and genralized to new relations in the latest version. For example, *complm* is genralized to *mark* and *infmod & partmod* to *vmod*. These relations were only considered by [12].

**Table 39** TDs processed to extract the artefacts

| TDs Analyzed | [2] | [6] | [12] | [15] | [25] | iMER |
|---|---|---|---|---|---|---|
| nsubj(A, B) | ✓ | ✓ | ✓ | ✓ | ✓ | ✓ |
| nsubjpass(A, B) |  | ✓ | ✓ |  |  | ✓ |
| dobj (A, B) | ✓ | ✓ | ✓ | ✓ | ✓ | ✓ |
| iobj(A, B) |  |  | ✓ |  |  | ✓ |
| pobj(A, B) | ✓ |  | ✓ |  |  | ✓ |
| nmod:of(A,B) |  |  | ✓ |  | ✓ | ✓ |
| nmod:in(A,B) |  |  | ✓ |  |  | ✓ |
| nmod:to(A,B) |  |  |  | ✓ |  | ✓ |
| nmod:for(A,B) |  |  |  | ✓ |  | ✓ |
| nmod:from(A,B) |  |  |  |  |  | ✓ |
| nmod:as(A,B) |  |  |  |  |  | ✓ |
| nmod:by(A,B) |  |  |  |  |  | ✓ |
| nmod:agent(A,B) |  | ✓ | ✓ |  |  | ✓ |
| nmod:with(A,B) |  |  |  |  |  | ✓ |
| nmod:poss(A,B) |  |  | ✓ |  |  | ✓ |
| amod(A,B) | ✓ |  | ✓ | ✓ | ✓ | ✓ |
| compound(A,B) | ✓ | ✓ | ✓ | ✓ |  | ✓ |
| conj:and(A,B) |  |  |  |  |  | ✓ |
| conj:or(A,B) |  |  |  |  |  | ✓ |
| nummod(A, B) |  |  | ✓ |  |  | ✓ |
| det(A,B) |  |  |  |  |  | ✓ |
| xcomp (A, B), |  |  | ✓ | ✓ |  | ✓ |
| neg(A,B) |  |  | ✓ |  |  | ✓ |
| advcl:if (A, B) |  |  |  |  |  | ✓ |
| mark(A, B) |  |  | ✓ |  |  | ✓ |
| advmod(A, then) |  |  | ✓ |  | ✓ | ✓ |
| advmod(A, else) |  |  | ✓ |  |  | ✓ |
| nobjpass (A, B) |  | ✓ |  |  |  |  |
| cop(A, B) |  |  | ✓ |  |  |  |
| vmod(A,B) |  |  | ✓ | ✓ |  |  |

#### 5.5.2. Relations between TDs

Instead of procecssing the sentences individually, we analyzed the TDs by considering their relationship. Because it is possible that an element of an artefact is incomple in one TD, but by combing with others, we can get the complete information. Elements having compound names could not be generated from one TD independently. Therefore, we need to consider its surounding TDs. In our rules, it is stated that in addition to the independent processing, if *compound(A,B)* appears before other TDs, then we have to consider relation between them by matching the partcipating words. For example, by analyzing *compound(card-9, credit-8) & nsubj(enter-6, card-9),* we can extract the input entity "*credit card*".

### 5.5.3. Relationsships

In most of the exiting approaches, relationships between entities were extracted from transitive and other forms of the verbs. Due to the different stuctures of senetnces, there might be no transitive (diect relationship) between the entities. Therefore, we considered some additonal relations between the entities that include;

- *Prepositional relations*: In iMER, prepositional relations between the entities also consider, if there is no direct relationship. For instance, by processing *compound (card-5, credit-4) & nmod:of(card-5, customer-7),* we can extract two entities "*Credit card*" & "*Customer*" and relationship from the preposition "*of*". We converted these relations to a meaningful way of representation. So, this relationship could be represented as; "*Customer (has) Credit card*".

- *From the flow of data*: It is another way to extract the relationships. If no relation identified between the entities and these are generated from different sentences, then we have to find the flow of data between them. For instance, from "*A customer selects the date*" & "*System displays the available booking dates*". Two entities "*Customer*" & "*Booking*" are extracted but with no relation. However, customer is selecting "*date*" for booking so, we can present a relationship between them as "*Customer (selects) Booking.*"

## 6. Conclusion and Future Work

Conceptual model extraction is an important step in the SDLC. In this paper, we proposed an automated iterative approach (iMER) to extract the artefacts, i.e., ER and BP models from the different formats of requirements. A tool is developed using the Stanford CoreNLP 3.8 API for the pre-processing (tokenizing, lemmatizing, POS tagging and type dependencies generating) of the requirements documents. The key advantages of this technique are;

- It could process requirement documents in various formats, including general requirements, UCSs and User Stories.
- Iterative analysis of TDs for the extraction of conceptual models, resulting in more accurate results.
- Processed pronouns by replacing them with their anaphors.
- Entity relationship extraction that is typically restricted to a single sentence. iMER also extracted the relationships by examining the data flow in consecutive sentences.
- Cardinalities extraction with and without key indicators.
- BP model generated by considering the behavior of the requirements and interaction of operations with the entities.

In future, we would like to extend iMER to generate other conceptual models required for the implementation. This will help the developer to get the quality products with less effort.

# Appendix A

*Appendix A.1*

Rules to extract entities and attributes.

**TDR1.**
if Dependency= nsubj(A, B)  OR nsubjpass(A, B)
    if A=VB|VBN and B=Noun and B≠Basic_Attrib then
        If prevTD = "compound" then
            Entity.add(compound(B) + Compound(A))
        else  Entity.add(nsubj(B))

**TDR2.**
if Dependency= (nsubj(A, B)  OR nsubjpass(A, B))
    if A=VB|VBN and B=Noun and B = Basic_Attrib then
        If prevTD = "compound" then
            Attributes.add(compound(B) + Compound(A))
        else  Attributes.add(nsubj(B))

**TDR3.**
if Dependencies= (dobj (A,B) OR iobj(A,B) OR pobj(A,B))
  if A=VB and B=Noun and B≠Basic_Attrib and prevTD≠"amod" and prevTD≠"advmod" and VB≠ "entered" | "inputted" | "saved" |"added" | "has" then
        if prevTD = "compound" then
            Entity.add(compound(B) + Compound(A))
        else  Entity.add(dobj(B))

**TDR4.**
if Dependencies= (dobj (A,B) OR iobj(A,B) OR pobj(A,B))
    if A=VB and B=Noun and (B = Basic_Attrib OR VB ="entered" | "inputted" | "saved" |"added" | "has") then
        If prevTD = "compound" then
            Attributes.add(compound(B) + Compound(A))
        else  Attributes.add(dobj(B))

**TDR5.**
if Dependencies= (dobj (A,B) OR iobj(A,B) OR pobj(A,B))
  if A=VB and B=Noun and (B = Basic_Attrib OR VB ="entered" | "inputted" | "saved" |"added" | "has") then
    If (prevTD = "amod" || prevTD = "advmod") and prev(B)="JJ" then
            Attributes.add(amod(B) + amod(A))
    else  Attributes.add(dobj(B))

**TDR6.**
if Dependency = nmod:of(A,B)
    if A=noun and B=Noun and A = Basic_Attrib and B≠Basic_Attrib then
        Entity.add(B)
        Attributes.add(A)
    if A=noun and B=Noun and A≠Basic_Attrib and  B≠Basic_Attrib then
        Entity.add(A)
        Entity.add(A)
    if A=noun and B=Noun and  A=Basic_Attrib and B=Basic_Attrib then
        Attributes.add(A + "of" + B)

**TDR7.**
if Dependency= nmod:in(A,B)
    if A=Noun and B=Noun then
        Entity.add(B)
        Attributes.add(A)

**TDR8:**
if Dependency= nmod:to(A,B) OR nmod:for(A,B) OR  nmod:from(A,B) OR nmod:as(A,B)
    if  B=Noun then
        Entity.add(B)

**TDR9.**
if Dependencies = nmod:by(A,B) OR nmod:agent(A,B) OR nmod:with(A,B)
    if B=Noun and B = Basic_Attrib then
               Attributes.add(B)
      else if B=Noun and B ≠ Basic_Attrib then
          Entity.add(B)

**TDR10.**
if Dependencies= nmod:poss(A,B)
  if A=Noun and B = Noun then
    Entity.add(B)
    Attributes.add(A)
  else if A=Noun and B= PREP ≠ Basic_Attrib then
    Attributes.add(B)

**TDR11.**
if Dependencies = amod(A,B)
  if A=Noun and B = JJ and A=basic_Attrib then
      Attributes.add(B + A)
  else if A=Noun and B=JJ and A ≠ Basic_Attrib and then
      Entity.add(A)

**TDR12.**
if Dependencies =compound(A,B) and nextTD≠nsubj and nextTD≠dobj
  if A=Noun and B = Noun and A=Basic_Attrib and B≠ Basic_Attrib then
       Attributes.add(B + A)
       Entity.add(B)
  else if A=Noun and B = Noun and A≠Basic_Attrib and B= Basic_Attrib then
       Attributes.add(A + B )
       Entity.add(A)
  else if A=Noun and B = Noun and A=Basic_Attrib and B= Basic_Attrib then
       Attributes.add(B+A)
  else if A=Noun and B = Noun and A≠Basic_Attrib and B≠ Basic_Attrib then
       Entity.add(B+A)
  EndIf

**TDR13.**
if Dependencies = nmod:and(A,B) OR nmod:or(A,B)
  if A=Noun and B = Noun and A=Basic_Attrib and B= Basic_Attrib then
      Attributes.add(A)
      Attributes.add(B)
  else if A=Noun and B = Noun and A≠Basic_Attrib and B≠ Basic_Attrib then
      Entity.add(A)
      Entity.add(B)

*Appendix A.2*
Rules to genrate relationships, carinalities and operations for flow

**TDR14.**
if nsubj(Verb, $E_1$) & dobj(Verb, $E_2$)
  relationship.add( $E_1$ (Verb) $E_2$)

**TDR15.**
if nsubjpass(VBN, $E_1$) and (nmod:agent (VBN, $E_2$) or nmod:by (VBN, $E_2$))
  relationship.add ( $E_1$ (VBN) $E_2$)

**TDR16.**
if nmod:of ($E_1$, $E_2$)
  relationship.add($E_1$ (has) $E_2$ )

**TDR17.**
if nsubj(Verb, $E_1$) and dobj(VBN, $E_2$) and nmod:of ($E_2$, $E_3$)
  relationship.add ($E_1$ (VB) $E_2$ )
  relationship.add ( $E_2$ ("has") $E_3$ )

**TDR18.**
if nsubj(VB, $E_1$) and dobj(VB, $E_2$) and nmod:to (VB, $E_3$)
  relationship.add ($E_1$ (VB) $E_2$)
  relationship.add ($E_2$ (VB+ "to") $E_3$ )
  relationship.add($E_1$ (VB+ "to") $E_3$)

**TDR19.**
if nsubjpass(VBN, $E_1$) and nmod:to (VBN, $E_2$)
  relationship.add ( $E_1$ (VBN + "to") $E_2$)

**TDR20.**
if nsubj(Verb, $E_1$) and nsubjpass(VBN, $E_2$) and nmod:to (VBN, $E_3$)
    relationship.add ($E_1$ (VB) $E_2$ )
    relationship.add,($E_1$ (VBN + "to") $E_3$ )
    relationship.add ( $E_2$ (VBN + "to") $E_3$)

**TDR21.**
if nsubj(VB, $E_1$) and nmod:in (VB, $E_2$)
  relationship.add ( $E_1$ (VB +"in") $E_2$)

**TDR22.**
if nsubj(VB, $E_1$) and nmod:for (VB, $E_2$)
  relationship.add ($E_1$ (VB + "for") $E_2$)

**TDR23.**
if nmod:as(VB, $E_1$) and dobj (VB, $E_2$)
  relationship.add ($E_1$ (VB) $E_2$)

*Appendix A.3*
Rules to genrate carinalities.

**TDR24.**
if dependency = amod( $E_1$, JJ)
 cardinalities.add ($E_1$ ">"JJ )

**TDR25.**
if dependency = nummod( $E_1$, CD)
 cardinalities.add ($E_1$ ">" CD )

**TDR26.**
if dependency = det( $E_1$, DT)
  if (DT="Each" OR "All" OR "some" OR "Any" OR "Many" OR "Every" OR "multiple")
    cardinalities.add ($E_1$ ">" N )
  if (DT= "a" OR "an" )
    cardinalities.add ($E_1$ ">" 1)

**TDR 27.**
if Dependencies= nsubj(A,B) OR nsubjpass(A,B) OR dobj (A,B) OR iobj(A,B) OR pobj(A,B) OR iobj(A,B) OR nmod:to(A,B) OR mark(A,B)
  if A=VB and A ∈ {input, enter, fill, click, select, add, record, process, validate} then
    while (TD≠nsubj || nsubjpass || dobj || iobj || pobj || mark)
      if(TD.B==attributes)
        Input_Data.add(B)
      End if
    End While
  End if
End if

**TDR 28.**
if Dependencies= nsubj(A,B) OR nsubjpass(A,B) OR dobj (A,B) OR iobj(A,B) OR pobj(A,B) OR iobj(A,B) OR nmod:to(A,B) OR mark(A,B)
  if A=VB and A ∈ {display, output, retrieve, show, view, print} then
    while (TD≠nsubj || nsubjpass || dobj || iobj || pobj || mark)
      if(TD.B==attributes)
        Output_Data.add(B)
      End if
    End While
  End if
End if

**TDR 29.**
if Dependencies= nsubj(A,B) OR nsubjpass(A,B) OR dobj (A,B) OR iobj(A,B) OR pobj(A,B) OR iobj(A,B) OR nmod:to(A,B) OR mark(A,B)
  if A=VB AND A ∈ {get, send, prepare } then
    while (TD≠nsubj || nsubjpass || dobj || iobj || pobj || mark)
      if(B='system')
        if(TD.B==attributes AND )
          Output_Data.add(B)
      Else if(B≠'system')
        if(TD.B==attributes AND )
          Input_Data.add(B)
    End While
  End if
End if

**TDR 30.**
if Dependencies= nmod:by(A,B) OR nmod:agent(A,B) OR nmod:with(A,B)
  if A=VB and A ∈ {inputted, entered, filled, clicked, selected, added, recorded, processed, validateed} then
    while (TD≠ nmod:by || nmod:agent || nmod:with)
      if(TD.B==attributes)
        Input_Data.add(B)
      End if
    End While
  End if
End if

**TDR 31.**
if Dependencies= nmod:by(A,B) OR nmod:agent(A,B) OR nmod:with(A,B)
  if A=VB and A ∈ {displayed, outputted, retrieved, showed, viewed, printed} then
    while (TD≠ nmod:by || nmod:agent || nmod:with)
      if(TD.B==attributes)
        Output_Data.add(B)
      End if
    End While
  End if
End if

**TDR 32.**
if Dependencies= nsubj(A,B) OR nmod:by (A,B)
  if A=VB and A ∈ {input, enter, fill, click, select, add, submit, choose} AND B=External Actor then
    User_Action.add(A)
  End if
  if A=VB and A ∈ {display, output, retrieve, show, view, print, calculate, process, update, delete, search. modify, edit, calculate, remove} AND B= System then
    System_Actions.add(A)
  End if
End if

**TDR 33.**
if Dependencies= nsubj(A,B) OR nmod:by (A,B)
  if A=VB and A ∈ {receive, accept, get, obtain, acquire, redeem}
    if B=System then
      User_Action.add(A)
    Else  User_Action.add(A)
    End if
  End if
End if

**TDR 34.**
if Dependencies= xcomp(A,B) OR amod(A,B) OR neg(A,B)
  if A ||B ∈ {error, fail, wrong, invalid, incorrect, not} then
    Exceptions.add(B + A)
  End if
End if

**TDR 35.**
if Dependencies= (advcl:if(A,B) OR mark(A,if))
  Susyem_Actions.add ("if" + dobj.B + dobj.A)
  while (TD≠advmod)
    if(TD.B==attributes)
      System_Actions.add(B)
    End if
  End While
End if
if Dependencies= advmod(A,then) and advmod(A,else)
  Susyem_Actions.add ("then" + dobj.B + dobj.A)
  while (TD≠advmod)
    if(TD.B==attributes)
      System_Actions.add(B)
    End if
  End While
End if
if Dependencies= and advmod(A,else)
  Susyem_Actions.add ("else" + dobj.B + dobj.A)
  while (TD≠NULL)
    if(TD.B==attributes)
      System_Actions.add(B)
    End if
  End While
End if

**TDR 36.**
if Dependencies= nsubj(A,B) and A="validate"
  Susyem_Actions.add (B + A)
  while (TD≠NULL)
    if(TD.B==attributes)
      System_Actions.add(B)
    End if
  End While
End if

**TDR 37.**
if Dependencies= nsubj(A,B) OR xcomp(A,B)
  if  A ∈ {continue, restart, go, repeat} then
    System_Action.add(A + nummod.B || dobj.B)
  End if
End if

# Appendix B
## Appendix B.1

**IEEE Template:** The following requirements are to explain the "Search" functionality.

1. Provide Search Facility
    1.1. The system shall enable user to enter the search text on the screen.
    1.2. The system shall enable user to select multiple options on the screen to search.
    1.3. The system shall display all the matching product Number,

**General Template:** Following requirements are related to the shopping cart.

FR01: The users shall be able to view the categories on the application's home page.
FR02: The users shall be able to view items in different categories.
FR03: The users shall be able add items to the cart.
FR04: The users shall be able to view more information about an item before adding it to the cart.

**Descriptive Template**: These requirements are to describe the library management system.

"A library issue loan items to customers.
Each customer is known as a member and is issued a membership card that shows a unique member number.
Along with the membership number, other details of a customer must be kept such as a name, address, and date of birth."

**Paragraph Template:** The following paragraph is to explain the "user registration" function.

"Customer can use it to register. It starts when the user want to register. Users are allowed to input the basic information such as username, password, confirm password, address, telephone number, e-mail address, postcode, real name and so on. If the information is correct, then user can finish registration."

## Appendix B.2

### UCS's Tamplate # 1

| | |
|---|---|
| **Name** | Search Product |
| **Actor** | Any customer |
| **Basic Flow** | This use-case starts when the user accesses the product search page |

1. The user accesses the search page and enters the search parameter
2. The system search based on the value and the chosen category and displays the list of products matching the criteria.
3. The user also has the option of searching the product by manufacturer's name (example, Microsoft etc).
4. Similarly the user has also an option for searching by categories. For example, the user can search in category like audio, video, portable, etc.
5. The user has the option of entering all the product details like the product name, manufacturer name, product category, etc. The user then clicks on the search button.

**Alternate Flow        Data Validation**
1. The user does not enter the search value and clicks on the search button. The system displays an error message saying that the user needs to provide a value for the search.
**No matches**
1. The system displays a message saying "Item Not Found".

| | |
|---|---|
| **Pre-conditions** | The user is logged in. |
| **Post-conditions** | The user has completed the search. |

### UCS's Template # 2

| | |
|---|---|
| **UseCaseID:** | Cancel Reservation |
| Goal In Context: | A customer wishes to cancel a reservation. |
| Pre-Condition: | A reservation has already been made. Actor has successfully navigated to the main options screen. |
| Success End Condition: | The selected reservation has been cancelled. |
| Failed End Condition: | The selected reservation has NOT been cancelled. |
| Trigger Event: | Selects the "Cancel Reservation" option. |

**Main Success Scenario:.**
1. Customer selects the "Cancel Reservation" option.
2. System displays a screen with an input field for a reservation number.
3. Customer enters a reservation number and clicks the "Submit" button.
4. If reservation number is valid the system will display the details of the reservation.

**Extensions:.**
2.a. Customer selects the "Cancel" option.
2.a.1. System displays the main options screen.
3.a. Customer selects the "Cancel" option.
3.a.1. System displays the main options screen.

**Business Rules:**
B1: Refund policy in-case of advance payment

## Appendix C

BP extracted from the second and third case studies are presented in this annexure.

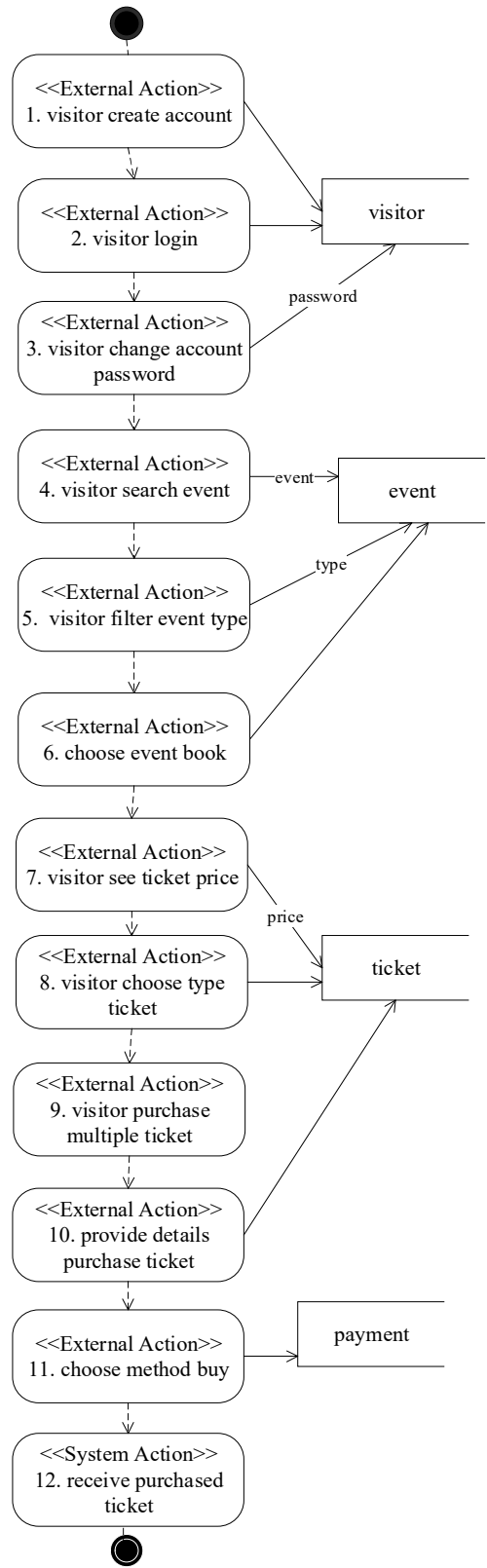

**Fig. 9.** Operations extracted from the User Stories stated in case study # 2

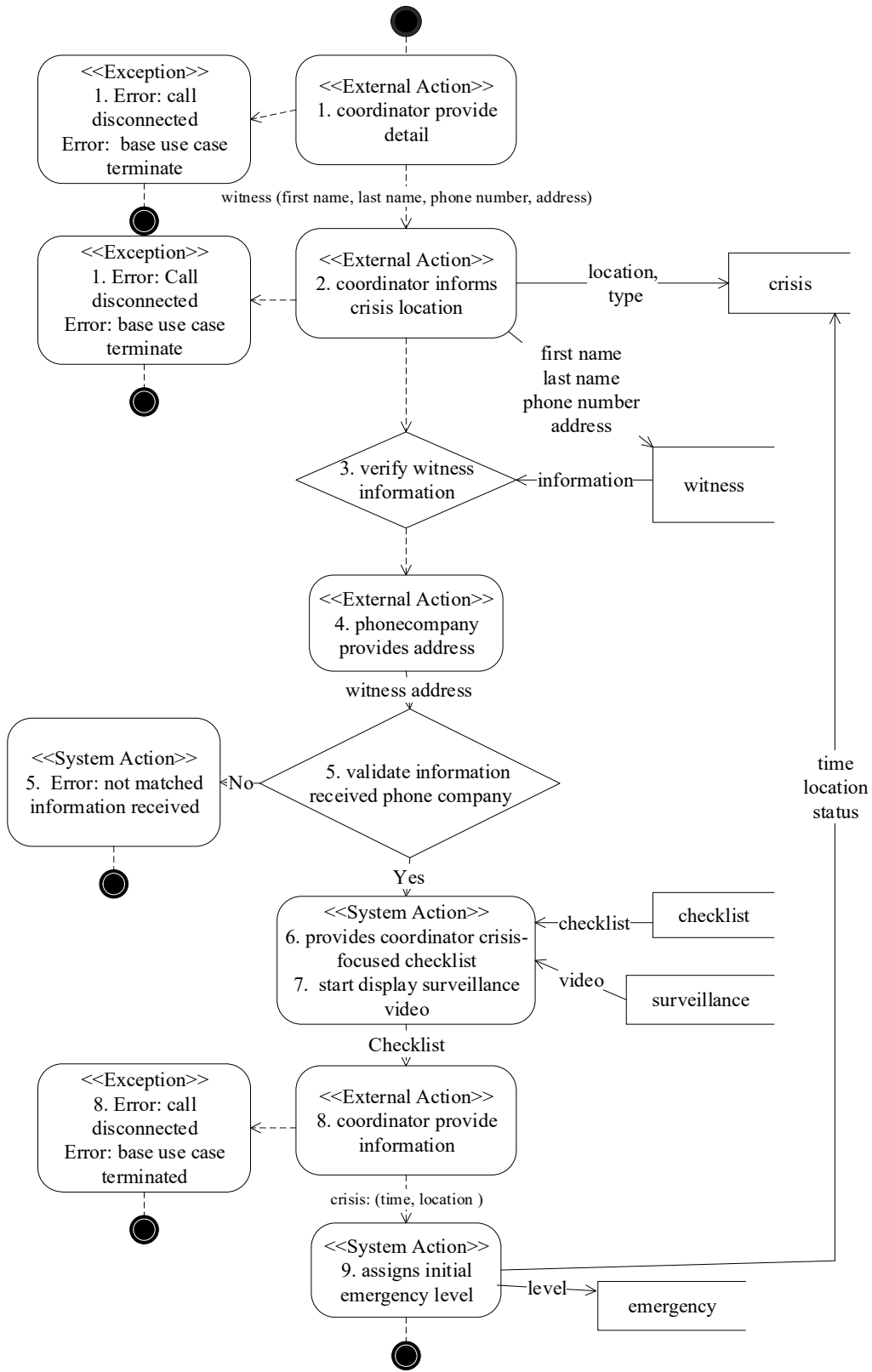

**Fig. 10.** BP model of "Capture Witness Report" UCS in case study # 3